\documentclass[preprint,twocolumn,5p]{elsarticle}

{}
{}
{}
\usepackage{amsfonts}

\usepackage{mathtools}
\usepackage{nccmath}
\usepackage{multirow}
\usepackage{graphicx} 
\usepackage{tabularx} 

\usepackage{color} 
\usepackage{lineno,hyperref}
\usepackage{subfigure}
\usepackage{amsmath}

\usepackage{lineno}

\journal{Renewable \& Sust. Energy Reviews \; {\url{https://doi.org/10.1016/j.rser.2019.109369}}}

\begin{document}
\nolinenumbers
\begin{frontmatter}
  \title{{\color{black}{Power systems with high renewable energy sources: a review of inertia and frequency control strategies over time}}}



\author[mymainaddress]{Ana Fern\'{a}ndez-Guillam\'{o}n\corref{mycorrespondingauthor}}
\ead{ana.fernandez@upct.es}


\author[mysecondaryaddress]{Emilio G\'{o}mez-L\'{a}zaro}
\ead{emilio.gomez@uclm.es}

\author[mytertiaryaddress]{Eduard Muljadi}
\ead{mze0018@auburn.edu}

\author[mymainaddress]{\'{A}ngel Molina-Garc\'{i}a}
\ead{angel.molina@upct.es}

\address[mymainaddress]{Dept. of Electrical Engineering, Universidad Polit\'{e}cnica de Cartagena, 30202 Cartagena, Spain}
\address[mysecondaryaddress]{Renewable Energy Research Institute and DIEEAC-EDII-AB, Universidad de Castilla-La Mancha, 02071 Albacete, Spain}
\address[mytertiaryaddress]{Dept. of Electrical and Computer Engineering, Auburn University, 220 Broun Hall, Auburn, AL 36849, USA}


\begin{abstract}
  Traditionally, inertia in power systems has been determined by
  considering all the rotating masses directly connected to the
  grid. During the last decade, the integration of renewable energy
  sources, mainly photovoltaic installations and wind power plants, has led to
  a significant dynamic characteristic change in power systems. This change is mainly due to the fact that most renewables have power electronics at the grid interface. 
  The overall impact on stability and reliability analysis of power systems is very significant. The power systems become more dynamic and require a new set of strategies modifying traditional generation control algorithms. Indeed, renewable generation units are decoupled from the grid by electronic converters, decreasing the overall inertia of the grid. `Hidden inertia', `synthetic inertia' or 'virtual inertia' are terms currently used to represent artificial inertia created by converter control of the renewable sources. Alternative spinning reserves are then needed in the new power system with high penetration renewables, where the lack of rotating masses directly connected to the grid must be emulated to maintain an acceptable power system reliability. {\color{black}{This paper reviews the inertia concept in terms of values and their evolution in the last decades, as well as the damping factor values. A comparison of the rotational grid inertia for traditional and current averaged generation mix scenarios is also carried out. 
  In addition, an extensive discussion on wind and photovoltaic power plants and their contributions to inertia in terms of frequency control strategies is included in the paper.}}
\end{abstract}


\end{frontmatter}
\nolinenumbers

\section*{Nomenclature}
\begin{table}[h]
	\begin{flushleft}
		\resizebox{0.875\columnwidth}{!}{ 
			\begin{tabular}{ll} 
				DFIG & Double Fed Induction Generator \\
				EU & European Union\\
				FSWT & Fixed Speed Wind Turbine \\
				HAWT & Horizontal Axis Wind Turbine\\
				PMSG & Permanent Magnet Synchronous Generator\\
				PV & Photovoltaic \\
				RES & Renewable energy sources \\
				ROCOF & Rate Of Change Of Frequency \\
				SCIG & Squirrel Cage Induction Generator \\
				VSWT & Variable Speed Wind Turbine \\
				WPP & Wind Power Plant\\
			\end{tabular}
		}
	\end{flushleft}
\end{table}

\section{Introduction} \label{sec.introduction}

Presently, power system stability relies on synchronous machines 
connected to the grid. They are synchronized to the grid and their stored kinetic energy 
is automatically extracted in response to a sudden power imbalance. For example, a sudden additional large load or a loss of a large generation unit from the grid, will slow down the machines on the grid and subsequently reduce grid frequency~\cite{d15}. However, the power systems generation fleet is changing from conventional generation to renewable energy sources (RES)~\cite{hadjipaschalis09}. Limited fossil fuel reserves and the importance of reducing greenhouse gases emissions are the main reasons for this transition in the electrical generation~\cite{bevrani10}. For instance, wind, solar and biomass generations overtook coal power in the EU for the first time during the year 2017~\cite{sandbag18}. However, some authors consider that 
only half of the overall electricity demand can be provided by RES~\cite{zakeri15,weitemeyer15}, despite the fact that it is expected that future electrical grids will be based on RES, distributed generation and power electronics~\cite{gross17on}. As an example, in Europe, it is expected that 323 and 192~GW of wind and PV will be installed in 2030, which will cover up to 30\% and 18\% of the demand, respectively~\cite{iea14,wind17}.

Among the different renewable sources available, PV and wind (especially doubly fed induction generators, DFIG~\cite{ochoa17}) are the two most promising resources for generating electrical energy~\cite{shah15}. 
Apart from their intermittency, 
they are connected through power converters which decouple them from the power system grid~\cite{tielens12,mohamed12}. Therefore, the effective inertia of the electrical grid is reduced when conventional generators are replaced by RES~\cite{erlich10,gautam11}, affecting the system stability and reliability~\cite{nguyen18}. This fact is considered as one of the main drawbacks of integrating a large amount of non-synchronous generators (i.e. RES) into the grid~\cite{du18}, as the frequency stability and its transient response is compromised~\cite{delille12}. Actually, low system inertia is related with $(i)$~a faster rate of change of frequency (ROCOF) and $(ii)$~larger frequency deviations (lower frequency nadir during frequency dips) within a short-time frame~\cite{daly15}.

In this work, we conduct an extensive literature review focusing on the inertia values for power systems
and wind power plants. 
The averaged inertia values
are estimated by different countries for the last two decades, by considering the 'effective' rotating
masses directly connected to the grid. In addition, the damping factor evolution is also included in the paper based on most of technical contributions and analysis found in the 
literature. {\color{black}The rest of the paper is organized as follows: inertia and damping factor analysis for power systems is discussed in detail in Section~{\ref{sec.inertia_analysis}}, determining the averaged inertia estimation for different countries; control strategies and contributions 
to integrate RES into grid frequency response is described in Section~{\ref{sec.stability}}
; finally, the conclusion is given in Section~\ref{sec:conclusion}.}

\section{Inertia analysis in power systems}\label{sec.inertia_analysis}
\subsection{Modeling the inertial response of a rotational synchronous generator: inertia constant analysis}\label{sec.inertial_response}

The group turbine-synchronous generator rotates due to two opposite torques: $(i)$ mechanical torque 
of the turbine, $T_{m}$ and $(ii)$ electromagnetic torque 
of the generator, $T_{e}$. The motion equation is~\cite{boldea15,fernandez17}:
\begin{equation}
\label{eq.j}
2H\dfrac{d\omega_{r}}{dt}=T_{m}-T_{e}\;,
\end{equation}
where both the $T_{m}$ and the $T_{e}$ are expressed in pu and $H$ the inertia constant in s.  
$H$ is given by:
\begin{equation}\label{eq.h}
H=\dfrac{1}{2}\cdot \dfrac{J\cdot\omega_{base}^{2}}{S_{base}},
\end{equation}
being $J$ the moment of inertia, $\omega_{base}$ the base frequency and $S_{base}$ the base power. $H$ determines the time interval during which the generator can supply its rated power only using the kinetic energy stored in the rotational masses of the generator. In Table~\ref{tab.H}, a review of $H$ values for different types of generation units and rated power is shown. 
\begin{table}[tpb]
	\begin{center}
		\resizebox{\columnwidth}{!}{ 
		\begin{tabular}{c c c c c } 
			\hline
			\bfseries Type of generating unit & \textbf{Rated power} &\bfseries $H$ (s) & \textbf{Reference} & \textbf{Year}\\ \hline
			Thermal & $500-1500$ MW & $2.3-2$ & \cite{anderson08} & 2008\\ 
			Thermal & 1000 MW & $4-5$ & \cite{dabur11} & 2011\\ 
			Thermal & 10 MW & 4 &\cite{de07} & 2007\\ 
			Thermal & Not indicated & $4-5$ & \cite{kumal12} & 2012\\ 
			Thermal (2 poles) & Not indicated & $2.5-6$ & \cite{kundur94} & 1994\\ 
			Thermal (4 poles) & Not indicated & $4-10$ & \cite{kundur94} & 1994\\  
			Thermal (steam) & 130 MW & 4 & \cite{tielens12} & 2012\\ 
			Thermal (steam) & 60 MW & $3.3$ & \cite{tielens12} & 2012\\ 
			Thermal (combined cycle) & 115 MW & $4.3$ & \cite{tielens12} & 2012\\ 
			Thermal (gas) & $90-120$ MW & 5 & \cite{tielens12} & 2012\\ 
			Thermal & Not indicated & $2-8$ & \cite{uned11} & 2011\\ 
			Hydroelectric $450<n<514$ rpm & $10-65$ MW & $2-4.3$& \cite{anderson08} & 2008\\ 
			Hydroelectric $200<n<400$ rpm & $10-75$ MW & $2-4$& \cite{anderson08} & 2008\\   
			Hydroelectric $138<n<180$ rpm & $10-90$ MW & $2-3.3$& \cite{anderson08} & 2008\\ 
			Hydroelectric $80<n<120$ rpm & $10-85$ MW & $1.75-3$& \cite{anderson08} & 2008\\ 
			Hydroelectric & Not indicated & $4,75$ & \cite{eremia13} & 2013\\ 
			Hydroelectric $n<$200 rpm& Not indicated & $2-3$ &\cite{grainger94} & 1994\\ 
			Hydroelectric $n>$200 rpm & Not indicated & $2-4$& \cite{grainger94} & 1994\\ 
			Hydroelectric & Not indicated & $2-4$ & \cite{kundur94} & 1994\\ \hline		
		\end{tabular}
	}
		\caption{Summary of inertia values ($H$) for different generation types.}
		\label{tab.H}
	\end{center}
\end{table}

Expressing Eq.~\eqref{eq.j} in terms of power, and considering the initial status as $0$, $P = P_{0}+\Delta P=(\omega_{r0}+\Delta\omega_{r})\cdot(T_{0}+\Delta T)$. For small deviations, the second order terms are neglected due to their small values, thus $\Delta P\simeq\omega_{r0}\cdot\Delta T+T_{0}\cdot\Delta \omega_{r}$, being $\Delta P=\Delta P_{m}-\Delta P_{e}$ and $\Delta T=\Delta T_{m}-\Delta T_{e}$. Furthermore, in steady-state $T_{m0}=T_{e0}$ and $\omega_{r0}=1$~pu. Hence, $\Delta P=\Delta P_{m}-\Delta P_{e}\simeq\Delta T_{m}-\Delta T_{e}$.

Therefore, if small variations around the steady-state conditions are considered, Eq.~\eqref{eq.j} can be written as Eq.~\eqref{eq.gp} in the time domain, or as Eq.~\eqref{eq.gp1} if the Laplace transform is applied.
\begin{equation}\label{eq.gp}
\dfrac{d\Delta\omega_{r}}{dt}=\dfrac{1}{2H}(\Delta P_{m}-\Delta P_{e})
\end{equation}
\begin{equation}\label{eq.gp1}
\Delta\omega_{r}=\dfrac{\Delta P_{m}-\Delta P_{e}}{2H\cdot s}
\end{equation}

Some loads (especially inverter-based loads) can also be modified to work as a load resource (demand response capability) under frequency deviations (e.g., motors driving compressors, pumps, industry loads, HVAC-heating ventilation air conditioning...). This fact can be modeled by including the damping factor $D$. As an example, for a synchronous machine, the electrical power $P_{e}$ can be then expressed as follows,
\begin{equation}\label{eq.Pe}
\Delta P_{e}=\Delta P_{L}+D\cdot\Delta\omega_{r}  ,
\end{equation}
where $P_{L}$ represents the load independent from frequency excursions.

\begin{figure}[tbp]
	\centering
	\includegraphics[width=0.95\columnwidth]{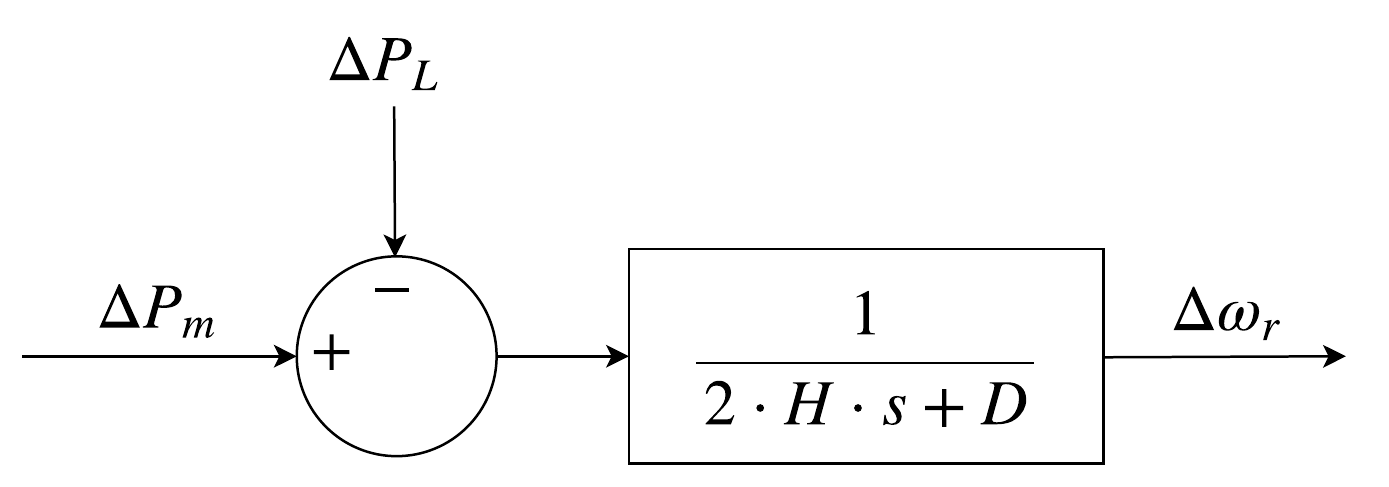}
	\caption{Block diagram representation of the swing equation}
	\label{fig.model}
\end{figure}

Substituting Eq.~\eqref{eq.Pe} into Eq.~\eqref{eq.gp1}, the mathematical representation of the motion of a synchronous generator is obtained. It is commonly referred to as \emph{swing equation}, see Eq.~\eqref{eq.sp}. It can be expressed in the form of a block diagram as shown in Figure~\ref{fig.model}. Hence, the initial response of a synchronous generator to a frequency event is governed by its stored kinetic energy at the rated frequency~\cite{spahic16}, 
\begin{equation}
\label{eq.sp}
\Delta\omega_{r}=\dfrac{\Delta P_{m}-\Delta P_{L}}{2H\cdot s+D}
\end{equation}

\subsection{Aggregated swing equation: equivalent inertia constant and damping factor analysis}\label{sec.agg_swing}

\begin{figure}[tbp]
	\centering
	\includegraphics[width=0.85\columnwidth]{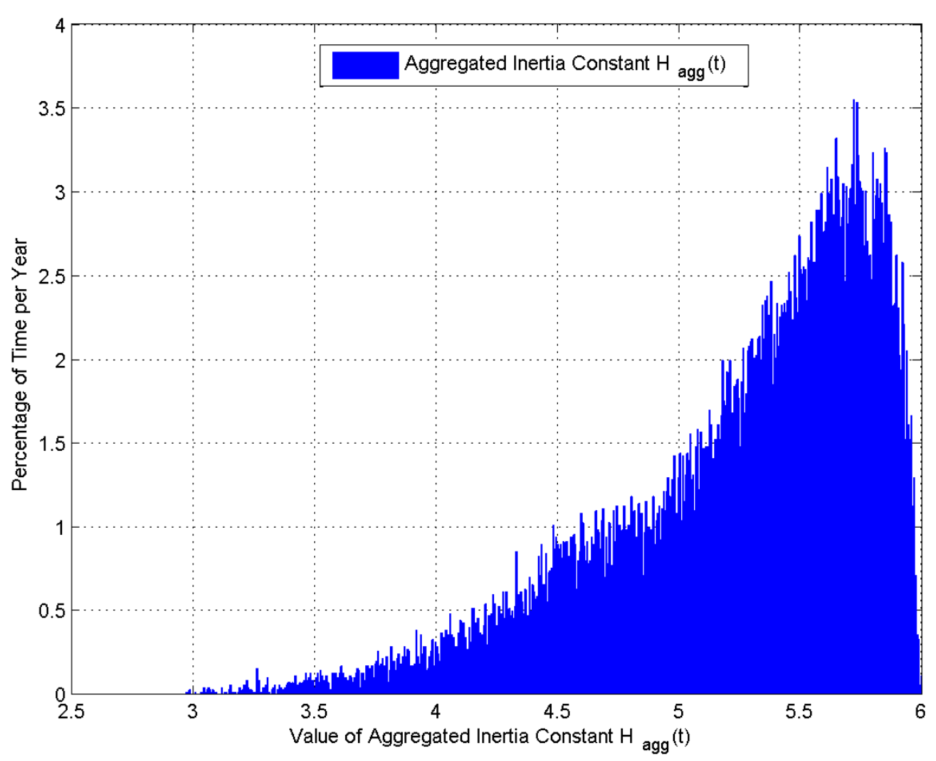}
	\caption{Histogram of equivalent inertia $H_{eq}$ in the German power system during 2012, \cite{ulbig14}}
	\label{fig.heq}
\end{figure}

\begin{figure}[tbp]
	\centering
	\includegraphics[width=0.85\columnwidth]{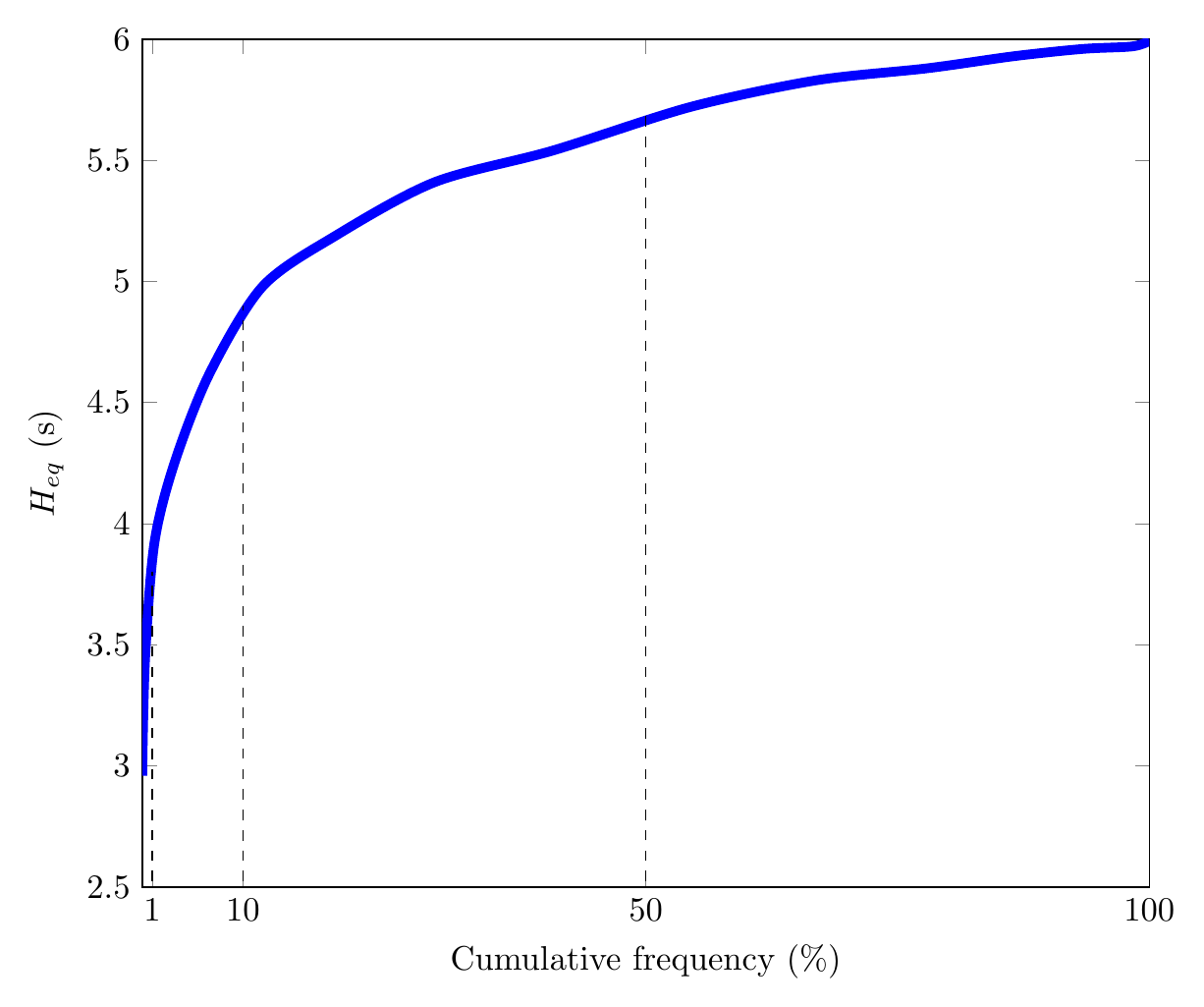}
	\caption{Cumulative frequency of the equivalent inertia $H_{eq}$ in the German power system during 2012}
	\label{fig.Hagg}
\end{figure}

{\color{black}In order to apply the swing equation to a power system, Eq.~\eqref{eq.sp} is rewritten. All synchronous generators are reduced to an equivalent rotating mass with an equivalent inertia $H_{eq}$,}
\begin{equation}
H_{eq}=\dfrac{\displaystyle\sum_{i=1}^{GCPS}H_{i}\cdot S_{base,i}}{S_{base}},
\label{eq.hagg}
\end{equation}
{\color{black} being $GCPS$ the number of generators coupled to the power system~\cite{chiodo18}, such as conventional power plants and FSWTs. In the past, it was considered that the equivalent inertial constant $H_{eq}$ of a power system was constant and time-independent. However, due to the RES integration and the variation in their generation throughout the day, the season of the year, etc., it is understood that $H_{eq}$ changes with time. An example of this variation is presented for the German power system during 2012 in~\cite{ulbig14}, see Figure~\ref{fig.heq}. From these data, the cumulative frequency curve is obtained and depicted in Figure~\ref{fig.Hagg}. It can be seen that during 50\% of the year 2012, the equivalent inertia was under 5.7~s; 10\% of the year, $H_{eq}$ was under 5~s; and only 1\% of the year, its value was under 4~s.}

{\color{black}In the same way as synchronous generators, all loads are
  grouped in an equivalent one with an equivalent damping factor
  $D_{eq}$. As stated in~\cite{huang13}, the impact of an inaccurate
  value of $D_{eq}$ is relatively small if the power system is stable,
  but this can be a major contribution under disturbances. Moreover,
  it is expected to decrease accordingly to the use of variable
  frequency drives~\cite{tielens16}.  Table~\ref{tab.damping}
  summarizes the different values proposed for the damping factor in the
  literature over recent decades.}

\begin{table}[tpb]
	\begin{center}
		\resizebox{\columnwidth}{!}{
			\begin{tabular}{c c c c } 
				\hline
				\textbf{Ref.} &  \textbf{Value ($pu_{MW}/pu_{Hz}$)} & \textbf{Analysis} & \textbf{Year}\\\hline
				\cite{kundur94} & 1--2 & Power system stability & 1994 \\ 
				\cite{ali11} & 0.83 & Two areas with non-reheat thermal units & 2011\\ 
				\cite{sudha11} & 1.66 & Two areas with thermal units & 2011 \\ 
				\cite{jiang12} & 1--1.8 & Three areas with non-reheat thermal units & 2012\\ 
				\cite{masuta12} & 2 & One area with nuclear, thermal, wind and PV & 2012\\ 
				\cite{shabani13} & 0.5 -- 0.9 & Three areas with non-linear thermal units & 2013 \\ 
				\cite{rout13} & 0.83 & Two areas non-reheat thermal units & 2013\\ 
				\cite{sahu13}& 0.83 & Two areas with thermal units & 2013 \\ 
				\cite{sathya15} & 0.83 & Two areas with reheat units & 2015 \\ 
				\cite{ruiz16} & 0.8 & IEEE 9 bus system with hydro-power, gas and wind turbines & 2016 \\ 
				\cite{dong17} & 1--1.8 & One and three areas with non-reheat thermal units & 2017 \\ 
				\cite{peng18} & 1--1.8 & Three areas with non-reheat thermal units & 2018 \\ 
				\cite{wu18} & 1 & Two areas with non-reheat thermal units & 2018 \\	\hline
			\end{tabular}
		}
		\caption{Damping factor values. Literature review}
		\label{tab.damping}
	\end{center}
\end{table}

{\color{black}By using Eq.~\eqref{eq.hagg}, an estimation of the
  equivalent inertia $H_{eq}$ of several parts of the world has been
  carried out by the authors. The International Energy Agency (IEA)
  provides global statistics about energy~\cite{iea}. By considering
  the annual averaged electricity, an averaged equivalent inertia
  constant ($H_{eq}$) provided by such conventional power plants
  ---Table~\ref{tab.H}--- can be estimated. Note that for this
  estimation, $S$ of Eq.~\eqref{eq.hagg} is replaced by the annual
  electricity value ($E_{g}$). The expression used to estimate the
  inertia is then Eq.~\eqref{eq.hestimated}, being $E_{g, total}$ the
  total electricity supplied (conventional+RES generation) within
  a year.}

\begin{equation}
\label{eq.hestimated}
H_{eq}=\dfrac{\displaystyle\sum_{i=1}^{GCPS}H_{i}\cdot E_{g,i}}{E_{g, total}}.
\end{equation}

\begin{figure}[tbp]
	\centering
	\subfigure[Generation mix in 1996]{\includegraphics[width=0.85\columnwidth]{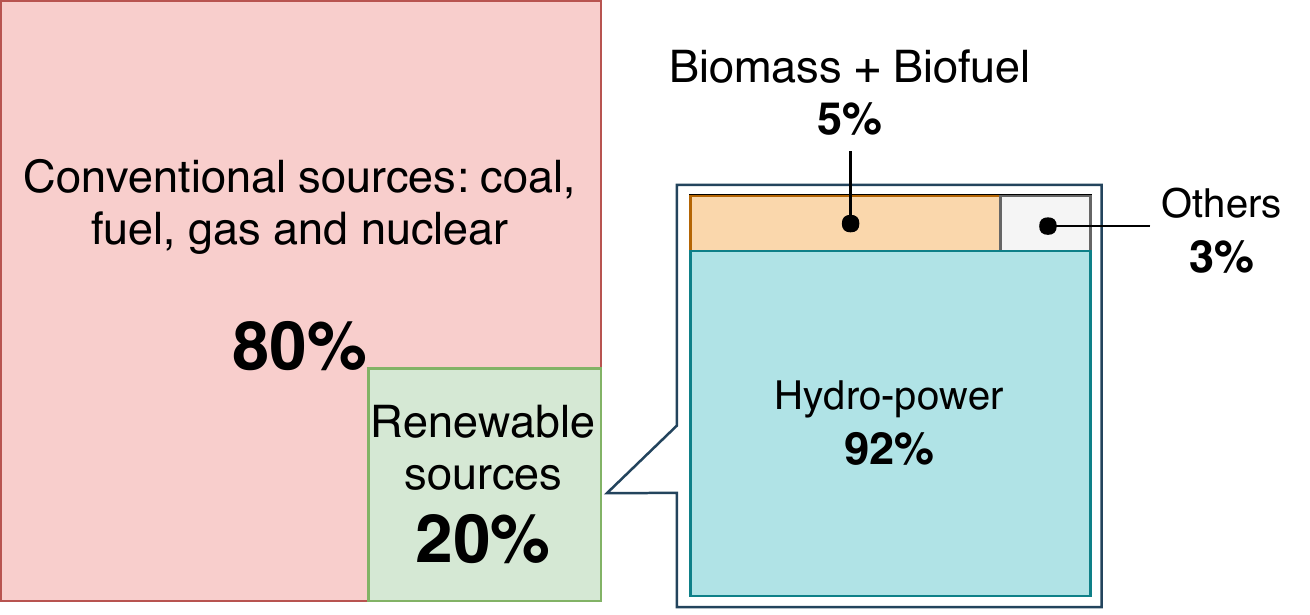}}
	\hfil
	\subfigure[Generation mix in 2016]{\includegraphics[width=0.85\columnwidth]{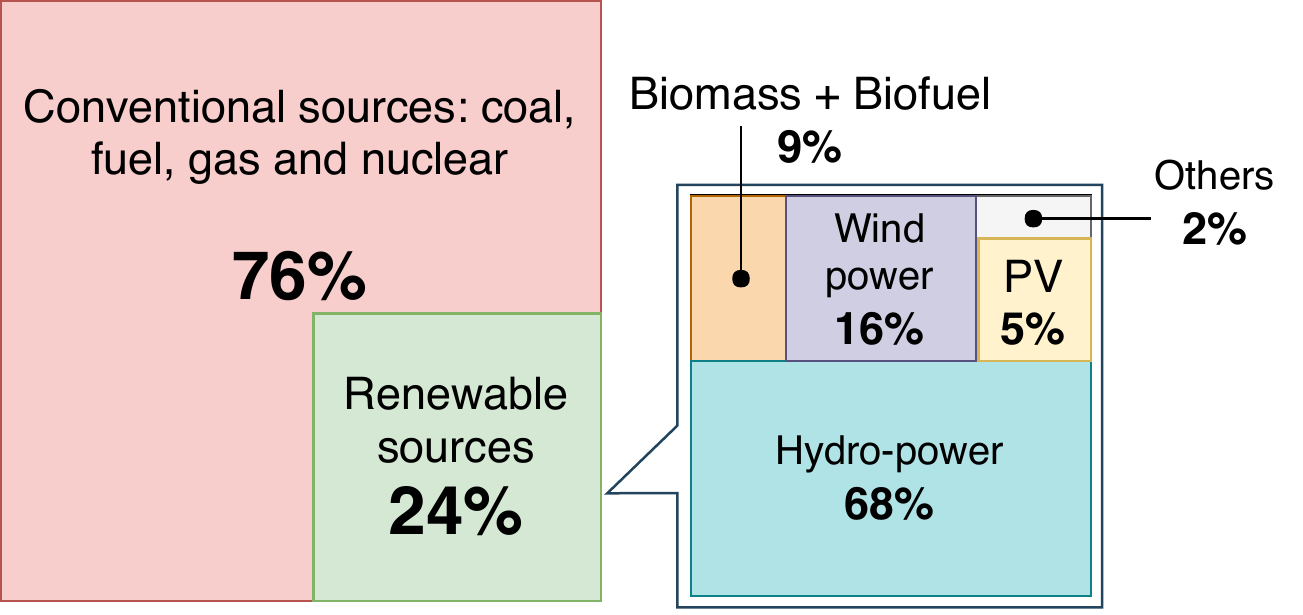}}
	\caption{Generation mix in the world: change between 1996 and 2016} 
	\label{fig.worldmix}
\end{figure}

\begin{figure}[tbp]
  \centering
  \includegraphics[width=0.9975\columnwidth]{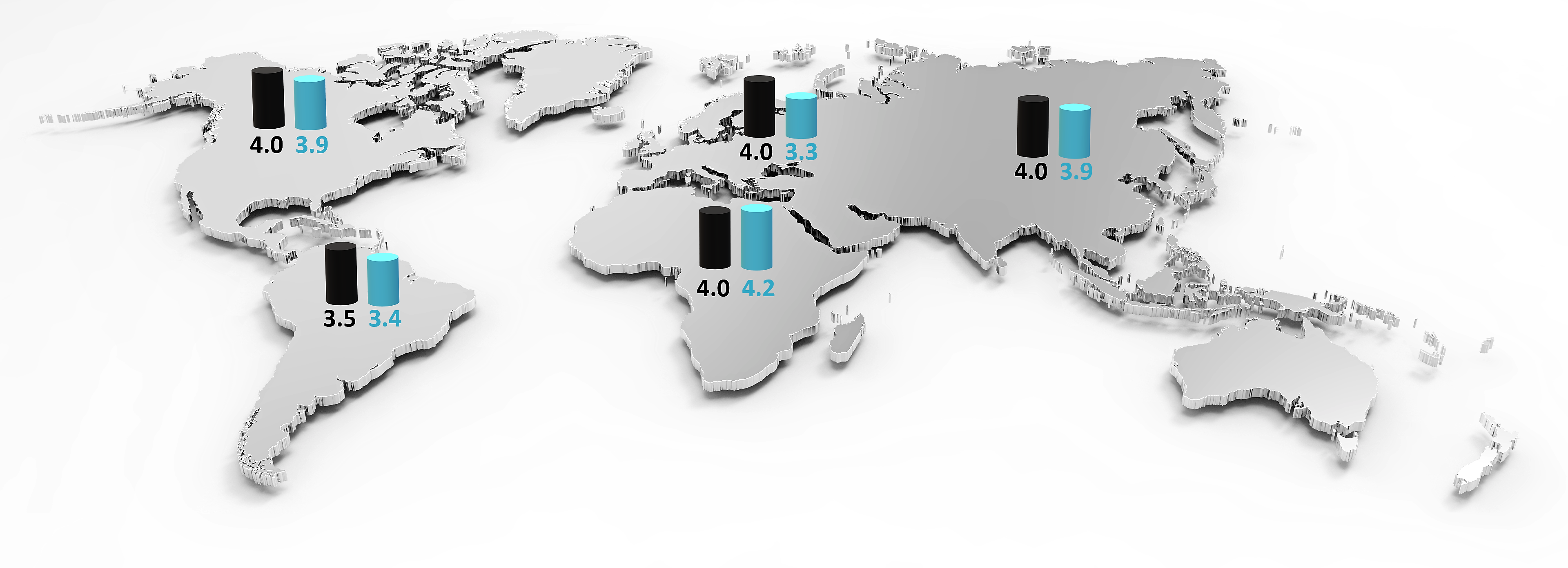}
  \caption{Equivalent inertia constants estimated in the world by continent. Change between 1996 and 2016} 
  \label{fig.world}
\end{figure}

{\color{black}Figure~\ref{fig.worldmix} shows a significant change in
  the averaged generation mix between 1996 and 2016. The total
  electricity consumption has been increased by more than 80\% within
  these two decades. However, RES generation has only increased by 4\% in
  the same two decades. Moreover, the share of the different renewable
  sources has changed significantly.  Indeed, the contribution share
  from hydro-power has been surpassed by biomass, biofuels, wind, and
  PV. Based on the approach previously described,
  Figure~\ref{fig.world} depicts the differences between the inertia
  constant for different continents in 1996 and in 2016. EU has
  reduced the equivalent inertia constant by nearly 20\%. In
  contrast, the reduction of inertia in Asia, USA, and South America
  lies between 2.5 and 3\%.}

\begin{figure}[tbp]
  \centering
  \includegraphics[width=0.85\columnwidth]{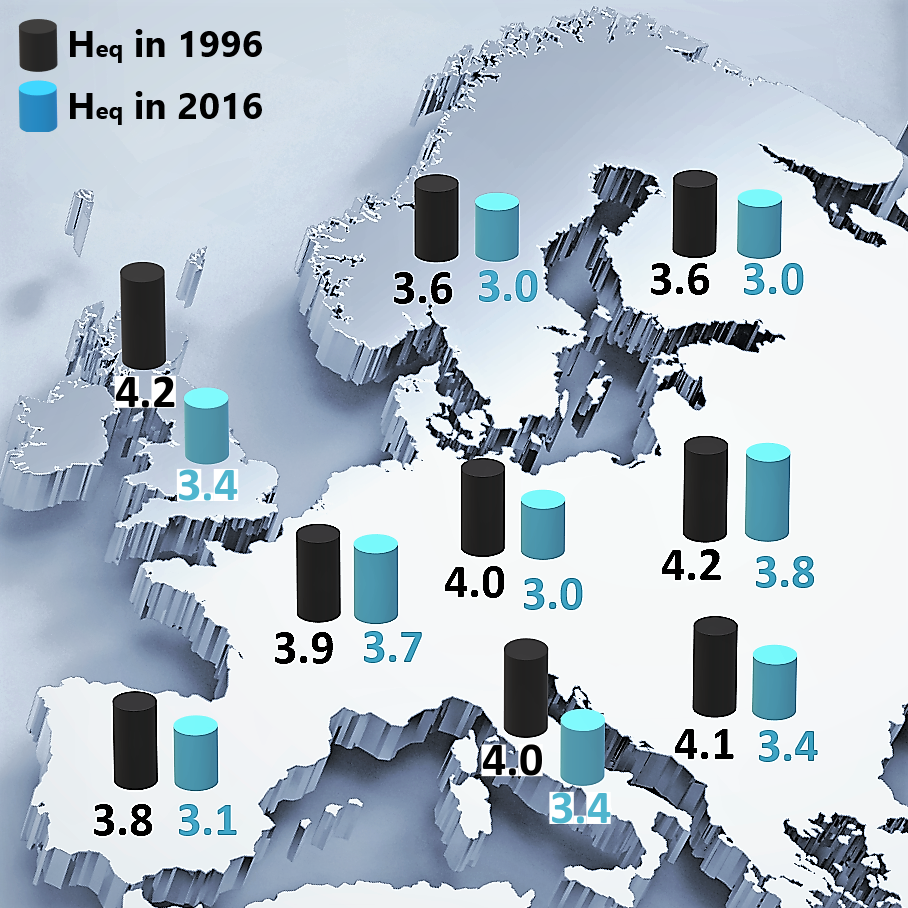}
  \caption{Equivalent inertia constants estimated in EU-28. Change between 1996 and 2016} 
  \label{fig.europe}
\end{figure}

\begin{figure}[tbp]
  \centering
  \includegraphics[width=0.99\columnwidth]{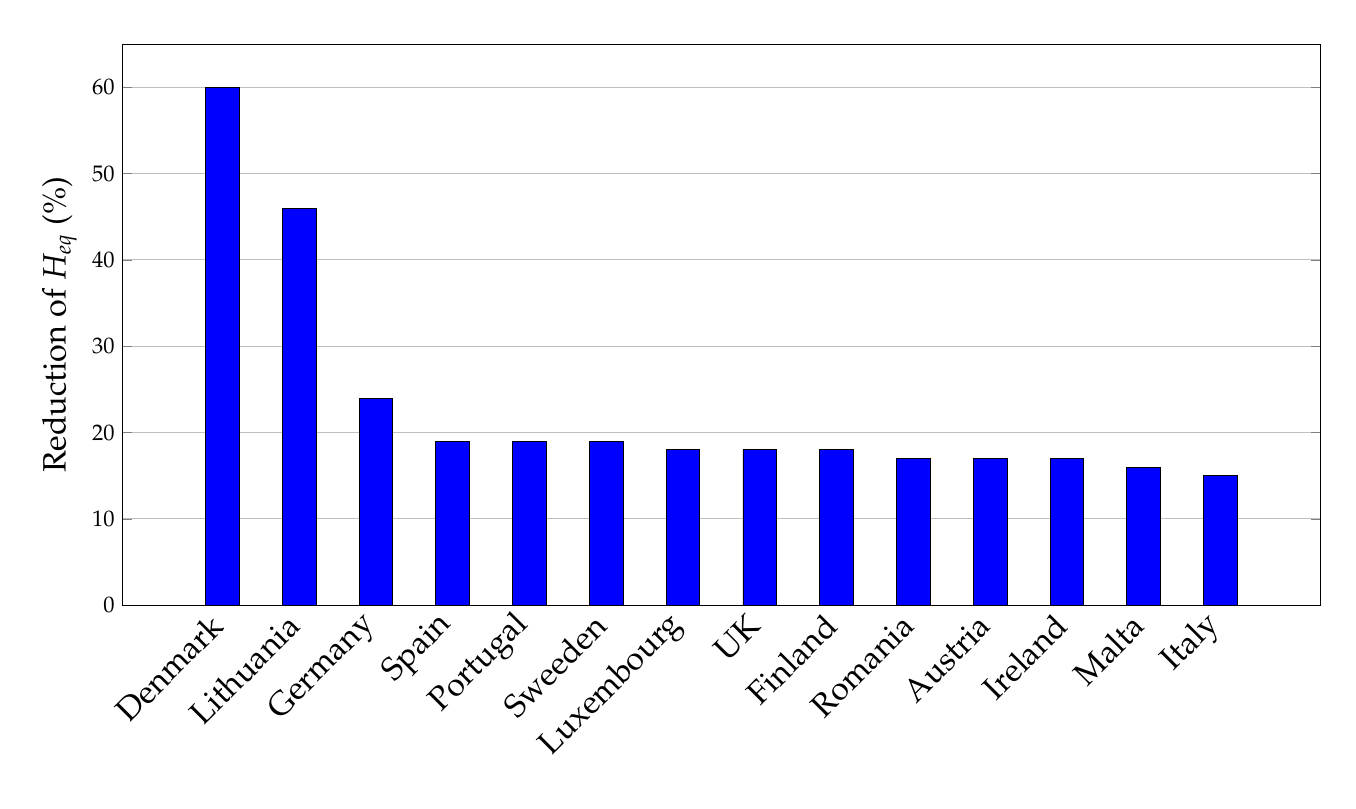}
  \caption{Equivalent inertia reduction in EU-28 between 1996 and 2016.} 
  \label{fig.heqs}
\end{figure}

\begin{figure}[tbp]
  \centering
  \includegraphics[width=0.99\columnwidth]{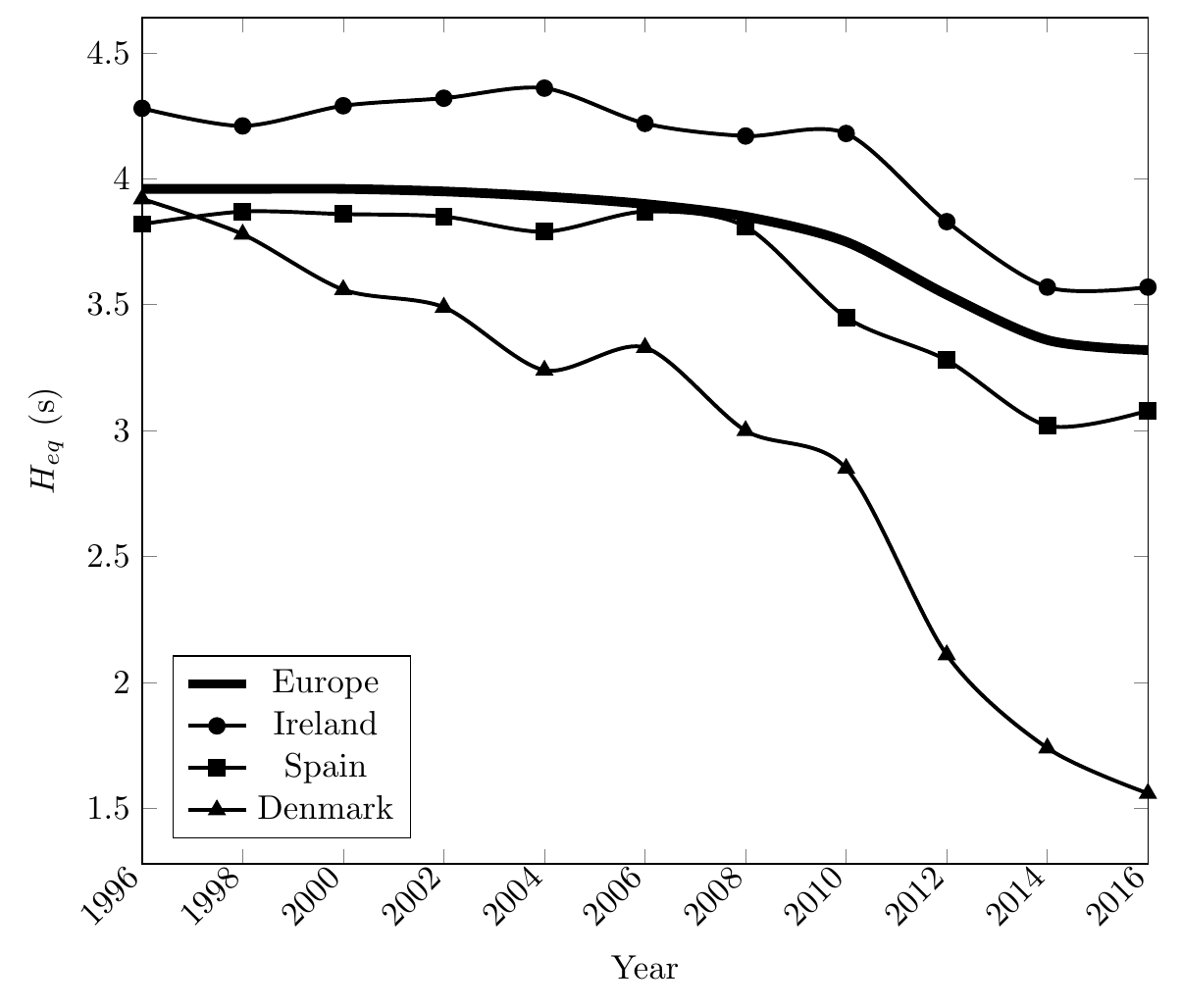}
  \caption{Evolution of equivalent inertia in EU-28 and some countries between 1996 and 2016.} 
  \label{fig.europe_evolution}
\end{figure}

\begin{figure}[tbp]
  \centering
  \subfigure[Generation mix in 1996]{\includegraphics[width=0.95\columnwidth]{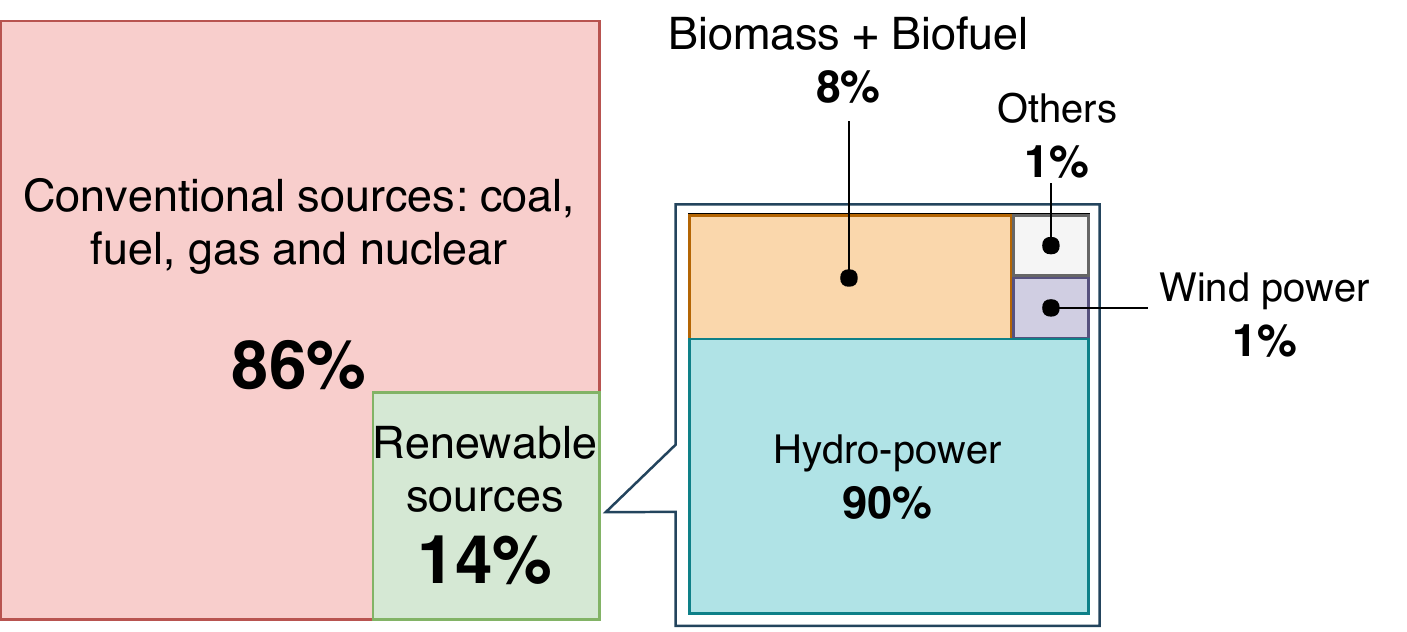}}
  \hfil
  \subfigure[Generation mix in 2016]{\includegraphics[width=0.95\columnwidth]{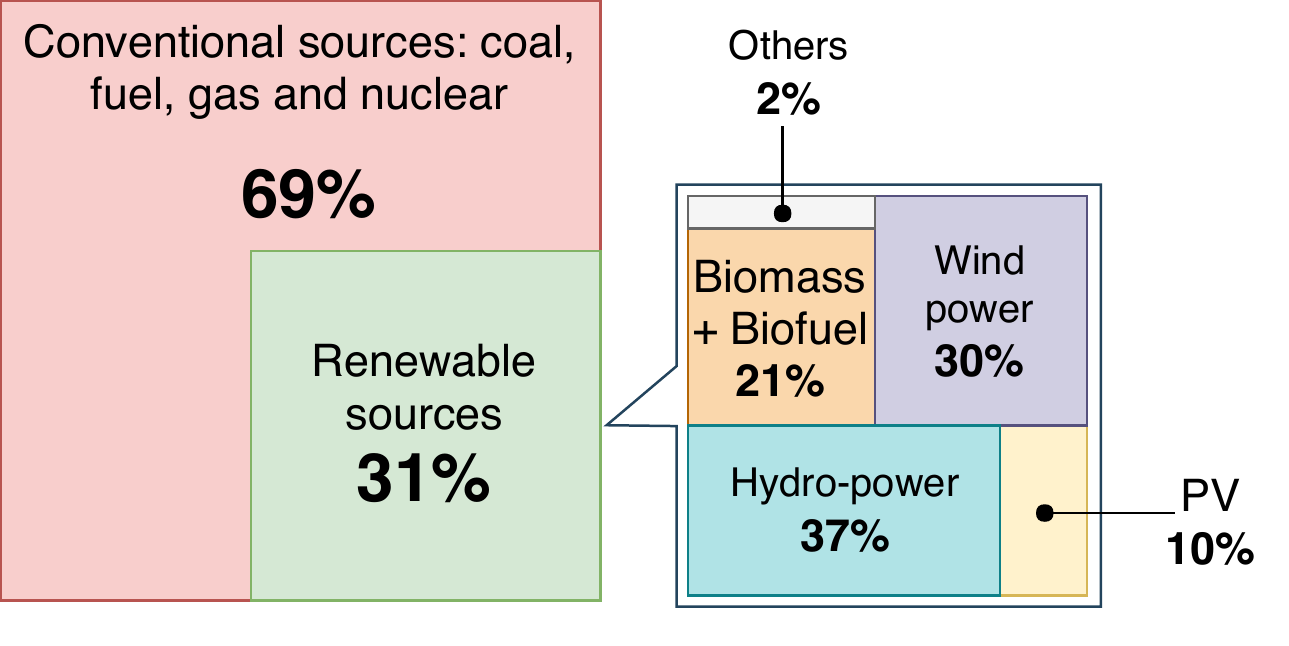}}
  \caption{Generation mix in Europe: change between 1996 and 2016.} 
  \label{fig.europemix}
\end{figure}

{\color{black}A more extensive analysis is conducted for the EU, where
  an average inertia reduction of 0.6~s can be estimated. In
  Figure~\ref{fig.europe}, an overview of the evolution of the
  equivalent inertia 
  in some EU countries is summarized. Similar information is 
  given in Figure~\ref{fig.heqs}, where the reduction of the
  equivalent inertia is 
  illustrated for those EU countries which have suffered a reduction
  larger than 15\% ($H_{eq}\;\mathrm{reduction} >
  15\%$). Figure~\ref{fig.europe_evolution} represents the equivalent
  inertia evolution of EU, as well as in three different countries
  (Ireland, Spain, and Denmark). For the EU, RES supply has increased
  nearly 
  by 20\%, in line with the 
  reduction of its inertia constant (refer to
  Figure~\ref{fig.europemix}). Similar to the generation mix in the
  world, wind, biomass, biofuels, and PV have 
  surpassed the development of hydro-power, which has
  drastically slowed down in recent years. }

\subsection{Modified equivalent inertia analysis: emulating hidden and virtual inertia from RES}\label{sec.modified_inertia}

{\color{black}To obtain the maximum power from the natural resource, both wind and PV power plants are controlled by power converters using the maximum power point tracking (MPPT) technique~\cite{muyeen10}. This power converter prevents wind and PV power plants to directly contribute to the inertia of the system, being thus referred to as 'decoupled' from the grid~\cite{zhao16}. As a consequence, to effectively integrate RES into the grid, frequency control strategies have been developed~\cite{gross17increasing,kerdphol17,teng17}. Such methods are commonly named as  synthetic, emulated or virtual inertia~\cite{vokony17}. If this emulation of inertia coming from RES was included in power systems, it would have to be considered to estimate the equivalent inertia. Then, this modified equivalent inertia would have two different components: $(i)$~synchronous inertia coming from conventional generators, $H_{S}$ and $(ii)$~emulated/virtual inertia coming from RES, $H_{EV}$~\cite{morren06phd,bevrani14,tielens16,gu17,tielens17}, modifying Eq.~\eqref{eq.hagg} to Eq.~\eqref{eq.hagg2}. $EVG$ is the number of RES connected to the grid through emulation/virtual control methods, and $H_{EV}$ is the inertia constant of the emulated/virtual generation unit.}

\begin{equation}
\label{eq.hagg2}
H_{eq}=\dfrac{\overbrace{\displaystyle\sum_{i=1}^{GCPS}H_{i}\cdot S_{base,i}}^{H_{S}}+\overbrace{\displaystyle\sum_{j=1}^{EVG}H_{EV,j}\cdot S_{base,j}}^{H_{EV}}}{S_{base}}.
\end{equation}

\begin{figure}[tbp]
  \centering
  \includegraphics[width=0.99\columnwidth]{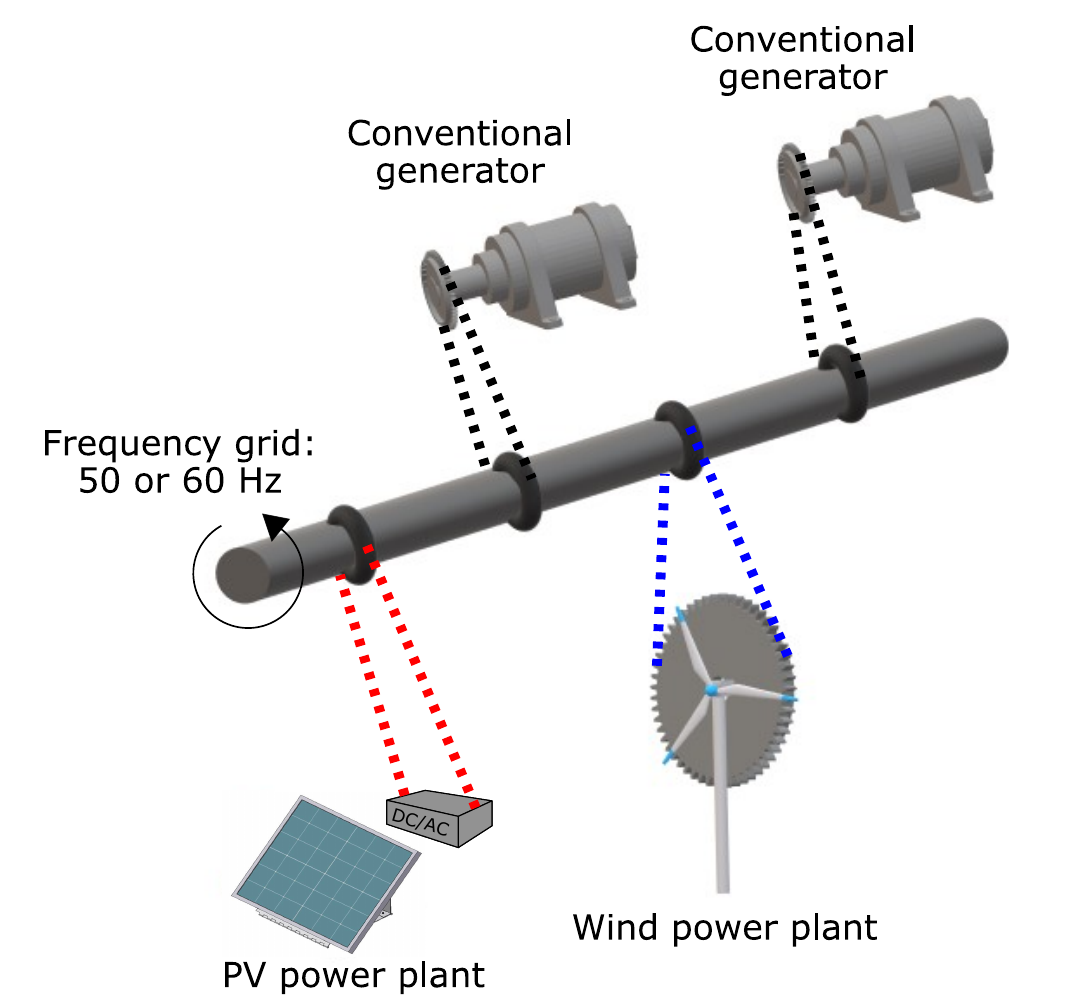}
  \caption{\color{black}Power system with synchronous, hidden and virtual inertia.}
  \label{fig.eps}
\end{figure}

\begin{figure}[tbp]
	\centering
	\includegraphics[width=0.95\columnwidth]{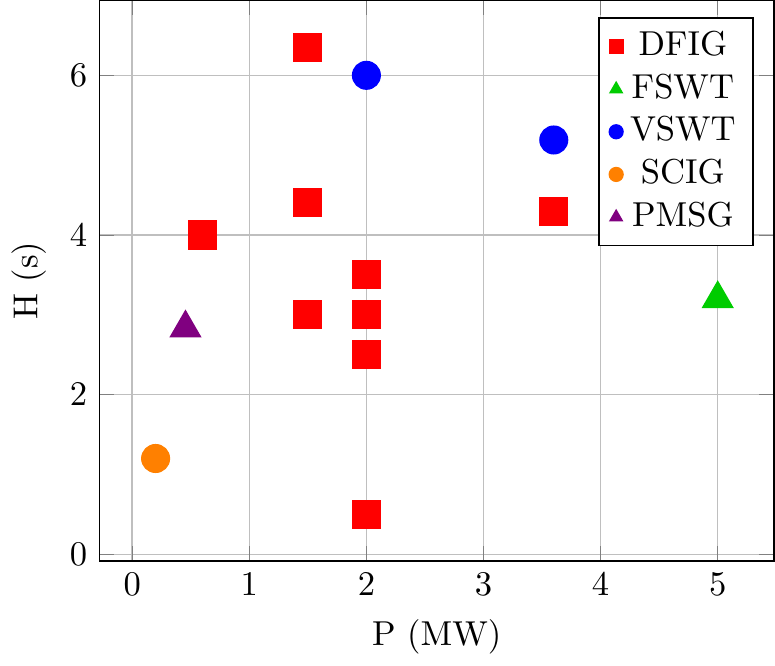}
	\caption{Inertia constant values ($H$) for different wind turbine technologies}
	\label{fig.dfig}
\end{figure}

This modified equivalent inertia expressed in Eq.~\eqref{eq.hagg2} is
graphically illustrated in Figure~\ref{fig.eps}, based
on~\cite{kroposki17}. {\color{black}Note the different representation
  between the coupling of VSWT and PV to the grid. The reason to this
  is that WPP has 'hidden' deployable inertia based on the kinetic
  energy stored in their blades, drive train and electrical
  generators, whereas PV has no stored kinetic energy due to the 
  absence of rotating masses.} 
{\color{black}Actually, modern VSWT have rotational inertia constants comparable to those of conventional generators~\cite{spahic16,thiesen16,simonetti18}. However, this inertia is 'hidden' from the power system point of view due to the converter~\cite{yingcheng11}. For instance, in Table~\ref{tab.Hwf} and Figure~\ref{fig.dfig}, the inertia constant of several types of wind turbines are summarized, and most of them are within the range $2-6$~s, in line with values presented for conventional units in Table~\ref{tab.H}. As a consequence, it is commonly considered that VSWT provide 'emulated hidden inertia', as rotational inertia could be provided by them~\cite{ruttledge15,van15,wang15,fischer16}. On the other hand, PV installations don't have any rotating masses~\cite{shah15,hosseinipour17}, having an inertia constant $H \approx 0$~\cite{tielens17phd}. Therefore, due to this absence of rotational masses and, subsequently, absence of inertia, the specific literature refers to the 'emulated synthetic/virtual inertia' provided by such PV power plants~\cite{nanou15,liu17,tang19,yang19}.}

\begin{table}[tpb]
	\begin{center}
		\resizebox{\columnwidth}{!}{ 
			\begin{tabular}{ c c c c c } 
				\hline
				\bfseries Type of wind turbine & \textbf{Rated power} &\bfseries $H$ (s) & \textbf{Reference} & \textbf{Year}\\ \hline
				Not indicated & Not indicated & $2-5$ & \cite{tielens12} & 2012\\ 
				Not indicated & 2 MW & $4.45$ & \cite{gonzalez07} & 2007\\  
				Not indicated & $2$ MW & $2.5$ & \cite{slootweg03} & 2003\\ 
				Not indicated & $16\cdot600$ kW & $3.7$ & \cite{salman03} & 2003\\ 
				HAWT with SCIG & $200$ kW & $1.2$ & \cite{kalantar10} & 2010 \\    
				FSWT & $10\cdot 500$ kW & $3.2$ & \cite{fox} & 2005\\  
				FSWT & Not indicated & $3.5$ & \cite{lalor05} & 2005 \\
				VSWT & 2 MW & 6 & \cite{morren06} & 2006\\ 
				VSWT & $3.6$ MW & $5.19$ & \cite{ullah08} & 2008 \\ 
				Types 1, 2, 3 & 1--5 MW & $2.4-6.8$ &\cite{ackermann05} & 2005\\ 
				DFIG & 2 MW & $3.5$ & \cite{ekanayake03} & 2003 \\ 
				DFIG & 660 kW & 4 & \cite{de06} & 2006\\ 
				DFIG & $1.5$ MW & $6.35$ & \cite{kayikcci09} & 2009\\ 
				DFIG & $1.5$ MW & $4.41$ & \cite{kayikcci09} & 2009\\ 
				DFIG & $3.6$ MW & $4.29$ & \cite{qu11} & 2011\\ 
				DFIG & $2$ MW & $3.5$ & \cite{holdsworth03} & 2003 \\  
				DFIG & $2$ MW & $2.5$ & \cite{perdana04} & 2004 \\ 
				DFIG & $660$ kW & 4 & \cite{de07} & 2007 \\ 
				DFIG (WPP) & $300$ MW & $1$ & \cite{xu07} & 2007 \\ 
				DFIG & $750$ MW & $5.4$ & \cite{gagnon05} & 2005 \\ 
				DFIG & $2$ MW & $3$ & \cite{arani13} & 2013 \\      
				DFIG & $1.5$ MW & $3$ & \cite{yang12} & 2012 \\     
				DFIG & $2$ MW & $0.5$ & \cite{xu06} & 2006\\        
				DFIG & $2$ MW & $3.5$ & \cite{ekanayake03comparison} & 2003 \\ 
				PMSG & $455$ kW & $2.833$ & \cite{westlake96} & 1996 \\\hline 
			\end{tabular}
		}
		\caption{Wind turbines inertia constants $H$ according to rated power and reference}
		\label{tab.Hwf}
	\end{center}
\end{table}

\begin{figure}[tbp]
	\centering
	\includegraphics[width=0.99\columnwidth]{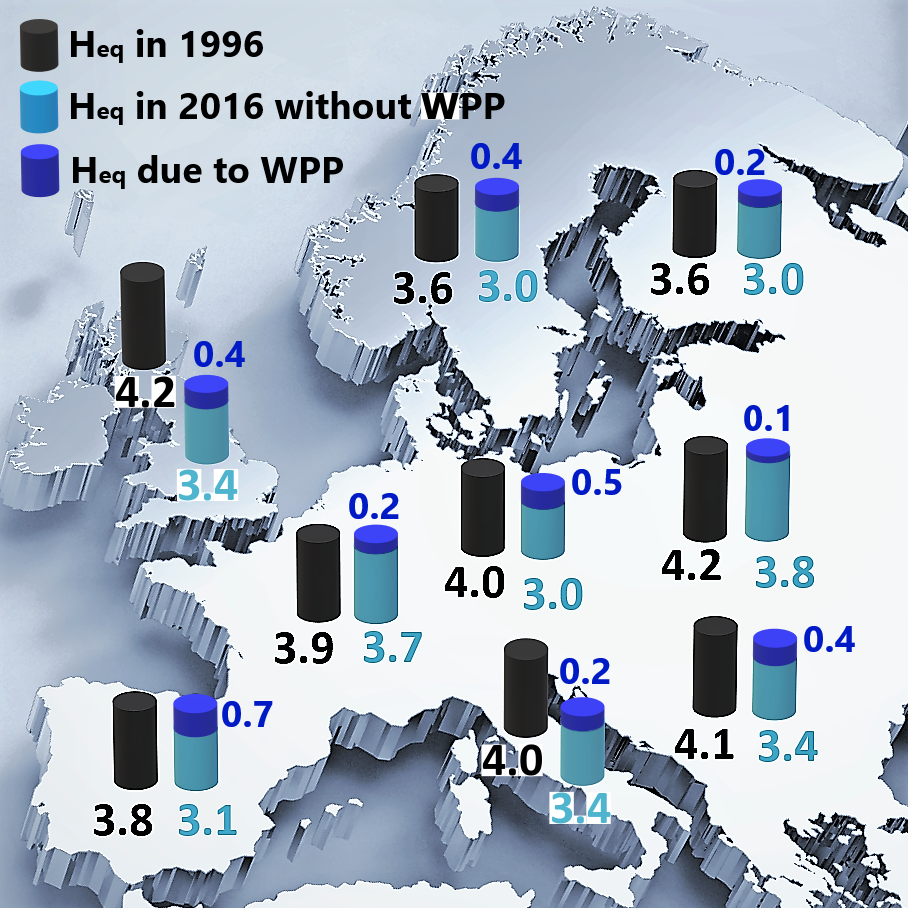}
	\caption{\color{black}Equivalent averaged inertia constants
          estimated in EU-28 considering emulated inertia provided by WPPs (1996--2016).}
	\label{fig.europe_mod}
\end{figure}

{\color{black}With regard to the equivalent inertia estimation for the
  EU, and considering the averaged hidden inertia of WPP depicted in
  Table~\ref{tab.Hwf}, the inertia change is reduced around 0.3~s,
  corresponding to 50\% of the value determined in
  Section~\ref{sec.agg_swing}. Figure~\ref{fig.europe_mod} presents
  the evolution of the equivalent inertia in the same EU countries of
  Figure~\ref{fig.europe}, being the dark blue values those due to the
  hidden inertia provided by VSWTs. As can be seen, by considering the
  hidden inertia of VSWT leads to a smaller reduction of the
  equivalent inertia.}


\section{RES frequency control strategies}\label{sec.stability}
\subsection{Preliminaries}

Generation and load in the power systems must be continuously balanced to maintain a steady frequency. Under any generation-load mismatch, grid frequency changes~\cite{rasolomampionona09}. Moreover, significant deviations from the nominal value may cause under/over frequency relay operations, and even lead to the disconnection of some loads from the grid~\cite{bevrani12}. Consequently, frequency stability is related to the ability of a power system to maintain the operating frequency close to its nominal value (i.e., 50 or 60~Hz, depending on the region) when an imbalance situation occurs~\cite{grigsby16}. Hence, frequency control is an essential component of a secure and robust electrical power system~\cite{ozer15}.

Frequency control is traditionally implemented by adjusting real power generation to balance the load. This traditional scheme has a hierarchical structure, and in Europe it is usually composed of three layers: primary, secondary and tertiary, from fast to slow timescales~\cite{zhao15}. The primary and secondary controls are automatic, while tertiary control is manually executed by the transmission system operator~\cite{ersdal16}. 

The primary frequency control (PFC) operates at a timescale up to low tens of seconds and uses a governor to adjust the mechanical power input around a set-point based on the local frequency deviation~\cite{zhao14}. It is the automatic response of the turbine governors 
in response to the deviations of the system frequency and depends on the setting of the speed-droop characteristics of each power plant~\cite{pandey13}. Therefore, each generating unit can be modeled with its speed governing system~\cite{li17designtest}. However, it does not restore grid frequency to its nominal value~\cite{guerrero09}. In Europe, primary control is triggered before the frequency deviation exceeds $\pm20$~mHz~\cite{entsoe}.

\begin{figure}[tbp]
	\centering
	\includegraphics[width=0.99\columnwidth]{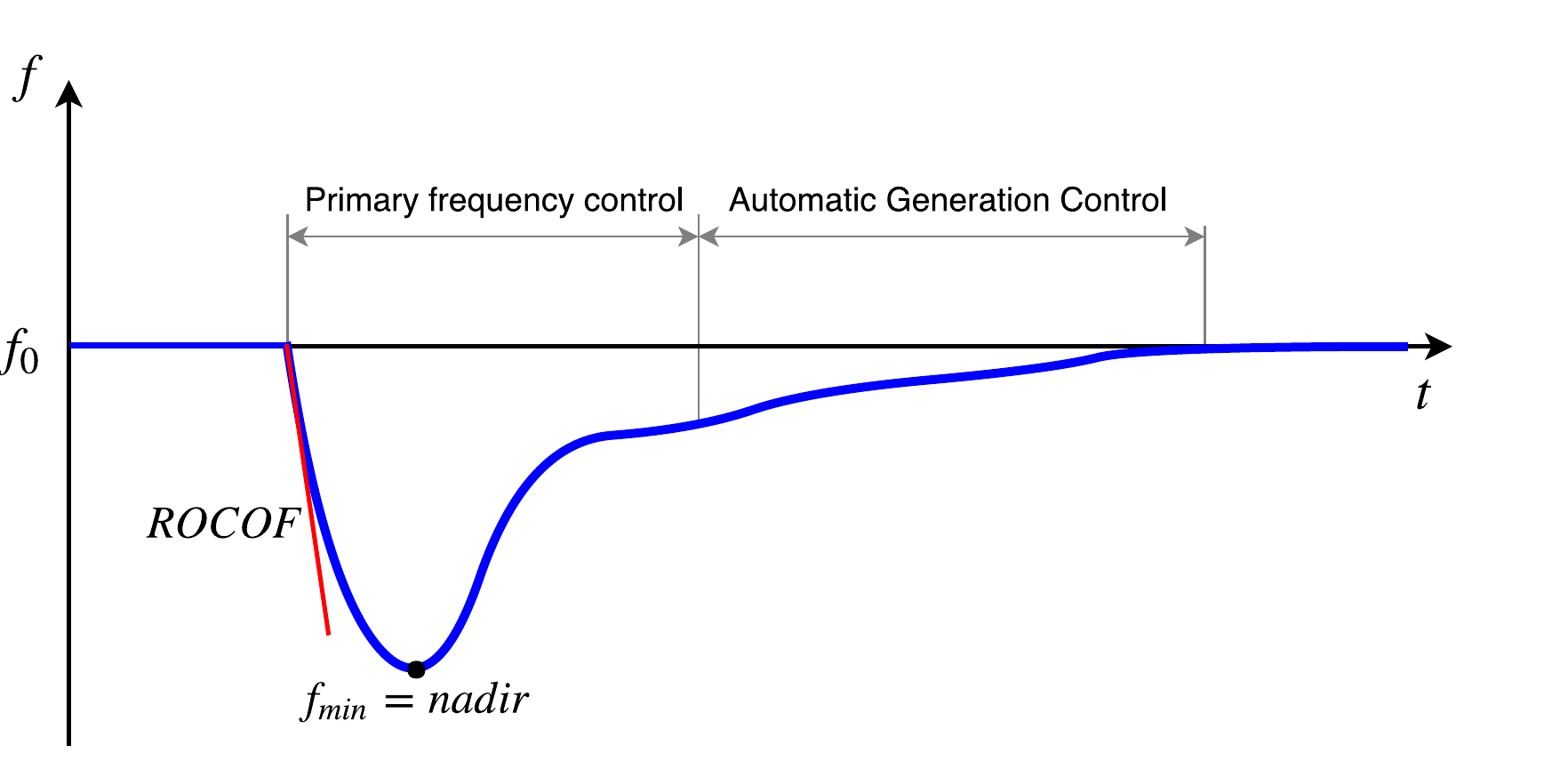}
	\caption{Frequency response after an imbalance} 
	\label{fig.frequency}
\end{figure}

Secondary frequency control or automatic generation control (AGC) removes the steady-state frequency deviation generated by the PFC~\cite{simpson15}. An integral controller modifies the turbine governor set-point to bring the frequency back to its nominal value~\cite{miao16}. It also keeps the scheduled exchanges between the different areas of an interconnected power system to their expected values~\cite{tan15}. In Europe, the time-frame is from seconds up to typically 15~min after an incident~\cite{entsoe}. Figure~\ref{fig.frequency} gives an example of a typical frequency excursion, where primary frequency control and AGC time intervals are shown.

Finally, the main objective of the tertiary frequency control is to perform an economically efficient generation-dispatch (economic dispatch)~\cite{perninge17}. Moreover, it is also intended to relieve transmission congestions and restoring the secondary control reserves~\cite{qureshi18}. This is also called security-constrained-economic dispatch (SCED).

\begin{figure}[tbp]
	\centering
	\includegraphics[width=0.99\columnwidth]{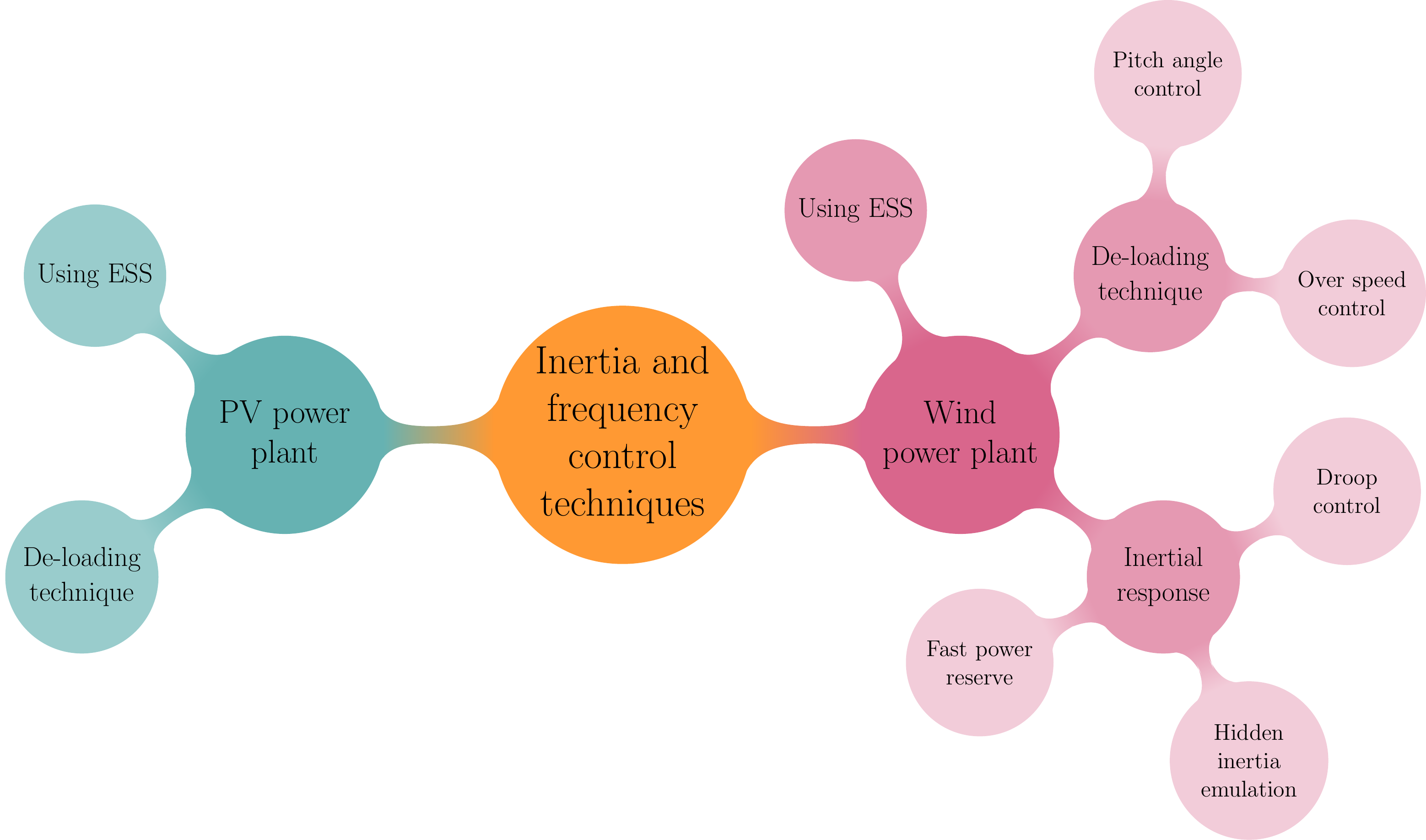}
	\caption{Inertia and frequency control techniques for RES}
	\label{fig.esquema}
\end{figure}

An increase in the penetration level of RES addresses a decreasing of the number of synchronous generators, leading to an initial decline in system inertia and power reserves for primary and secondary control~\cite{li17design}. Subsequently, low inertia is related to larger frequency deviations after a generation-load mismatch event~\cite{nedd17}, having implications on frequency related power systems dynamics \cite{ulbig15}. 
It is important to note that the rate of change of frequency (ROCOF) is strongly affected by 
the inertia available in the system~\cite{junyent15}. By this means, it is necessary that RES 
become an active role in grid frequency regulation, providing active power support under disturbances~\cite{you17}. The different technologies proposed to give additional inertia and frequency control from RES are usually classified as summarized in Figure~\ref{fig.esquema}.  

\subsection{PV power plant frequency control strategies}

PV power plants can use ESS such as batteries~\cite{marcos14,salim17,zhao18}, super-capacitors~\cite{taghizadeh15,you18} 
and flywheels~\cite{zhao18} in order to provide additional active power in an imbalanced situation.

\begin{figure}[tbp]
	\centering
	\subfigure[$V_{del,1}>V_{MPP}$]{\includegraphics[width=0.85\columnwidth]{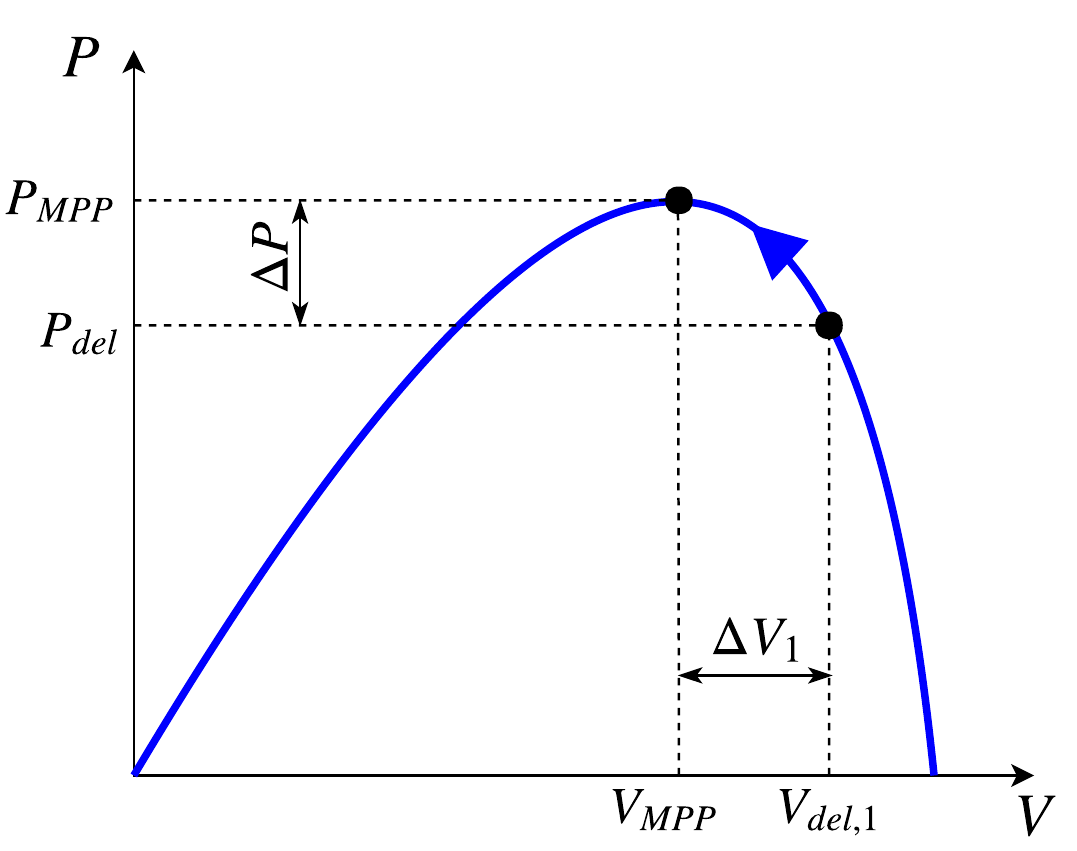}}
	\hfil
	\subfigure[$V_{del,2}<V_{MPP}$]{\includegraphics[width=0.85\columnwidth]{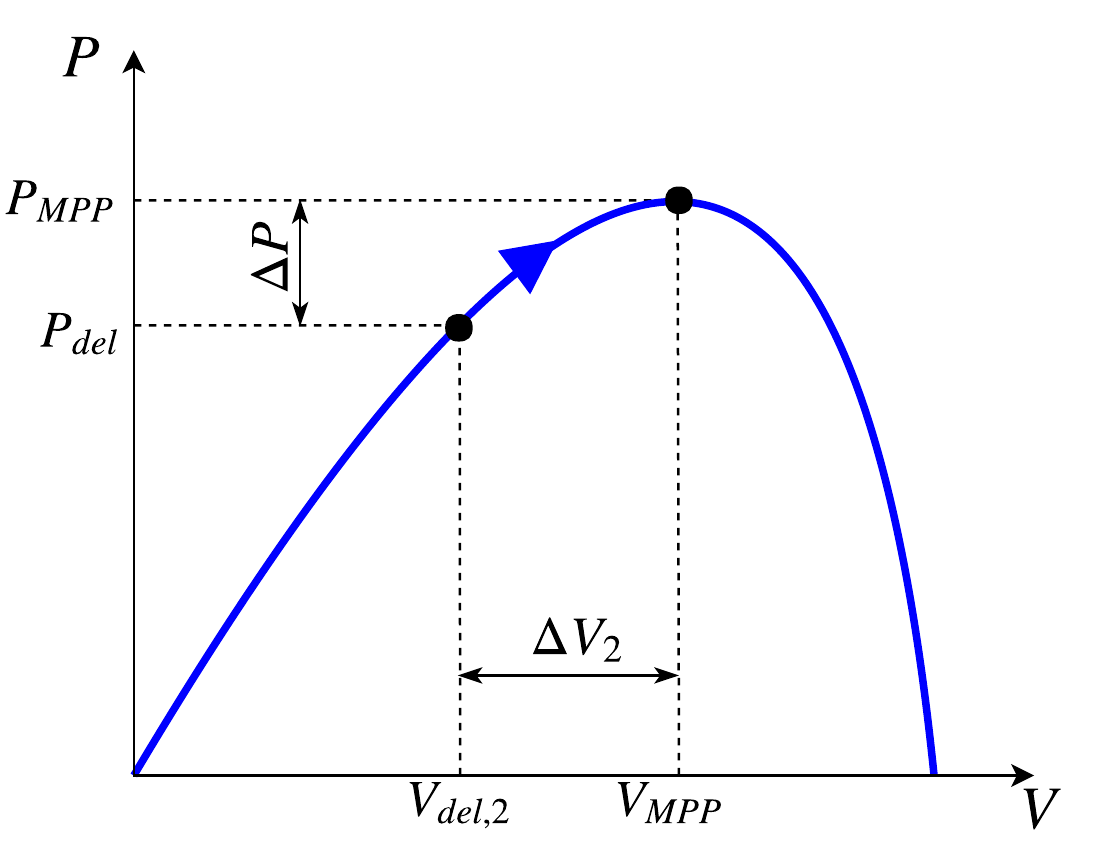}}
	\caption{Deloading techniques for PV}
	\label{fig.P-V_pv}
\end{figure}

A different strategy to be considered is the `de-loading technique' of the PV plant. It is based on operating these generating units below their optimal generation point, in order to have a certain amount (headroom) of active power to supply real power to the grid in case of a frequency-dip contingency~\cite{ziping17}. In general, PV power plants operate at the maximum power point tracking mode according to certain meteorological conditions (i.e., temperature $T$ and irradiation $G$), maximizing the 
revenues from selling energy~\cite{xin13}. Contributions focused on this technique can be found in~\cite{alatrash12,zarina12,zarina12photovoltaic,mishra13,rahmann14,zarina14}. By curtailment, we are operating the PV plant at a de-loaded point $P_{del}$, below $P_{MPP}$, so that the PV plants are able to support system frequency, as some power reserves $\Delta P=P_{MPP}-P_{del}$ are available. As depicted in Figure~\ref{fig.P-V_pv}, $P_{del}$ involves two different voltages: $(i)$~over the maximum power point voltage, $V_{del,1}>V_{MPP}$ and $(ii)$~under the maximum power point voltage, $V_{del,2}<V_{MPP}$. Due to stability concerns, the de-loaded voltage corresponds to the higher value~$V_{del,1}$~\cite{moutis15}. 

\subsection{Wind power plant frequency control strategies}

As in the PV power plants, wind power plants can also use ESS to provide additional power boost during an imbalanced situation (i.e., frequency dips). Batteries~\cite{salim17}, super-capacitors\cite{taghizadeh15,xiong18} and flywheels~\cite{jauch16} are proposed in the literature review.

\begin{figure}[tbp]
	\centering
	\includegraphics[width=0.9\columnwidth]{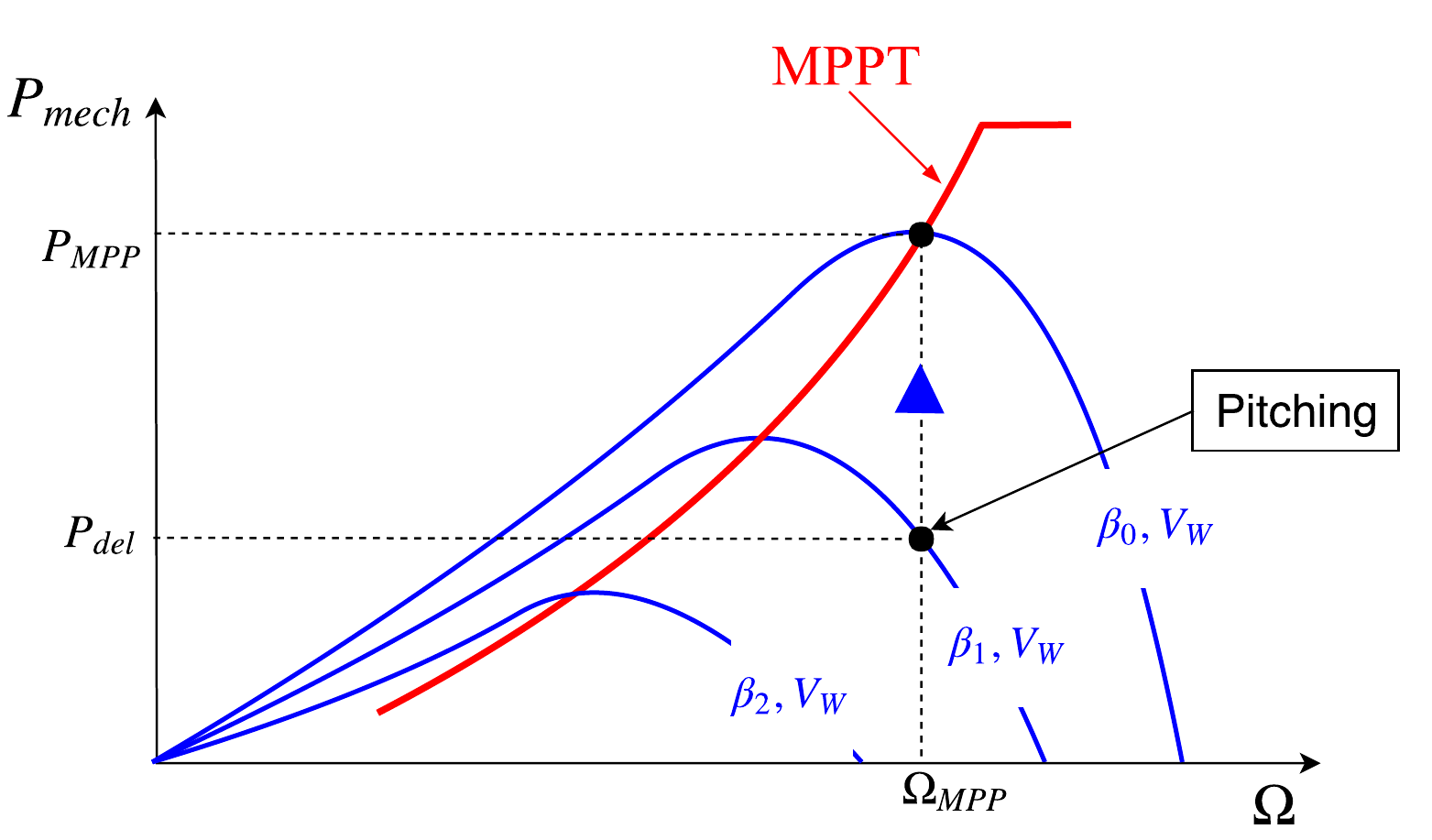}
	\caption{Pitch control}
	\label{fig.deload-wf}
\end{figure}

\begin{figure}[tbp]
	\centering
	\subfigure[Over-speed control]{\includegraphics[width=0.79\columnwidth]{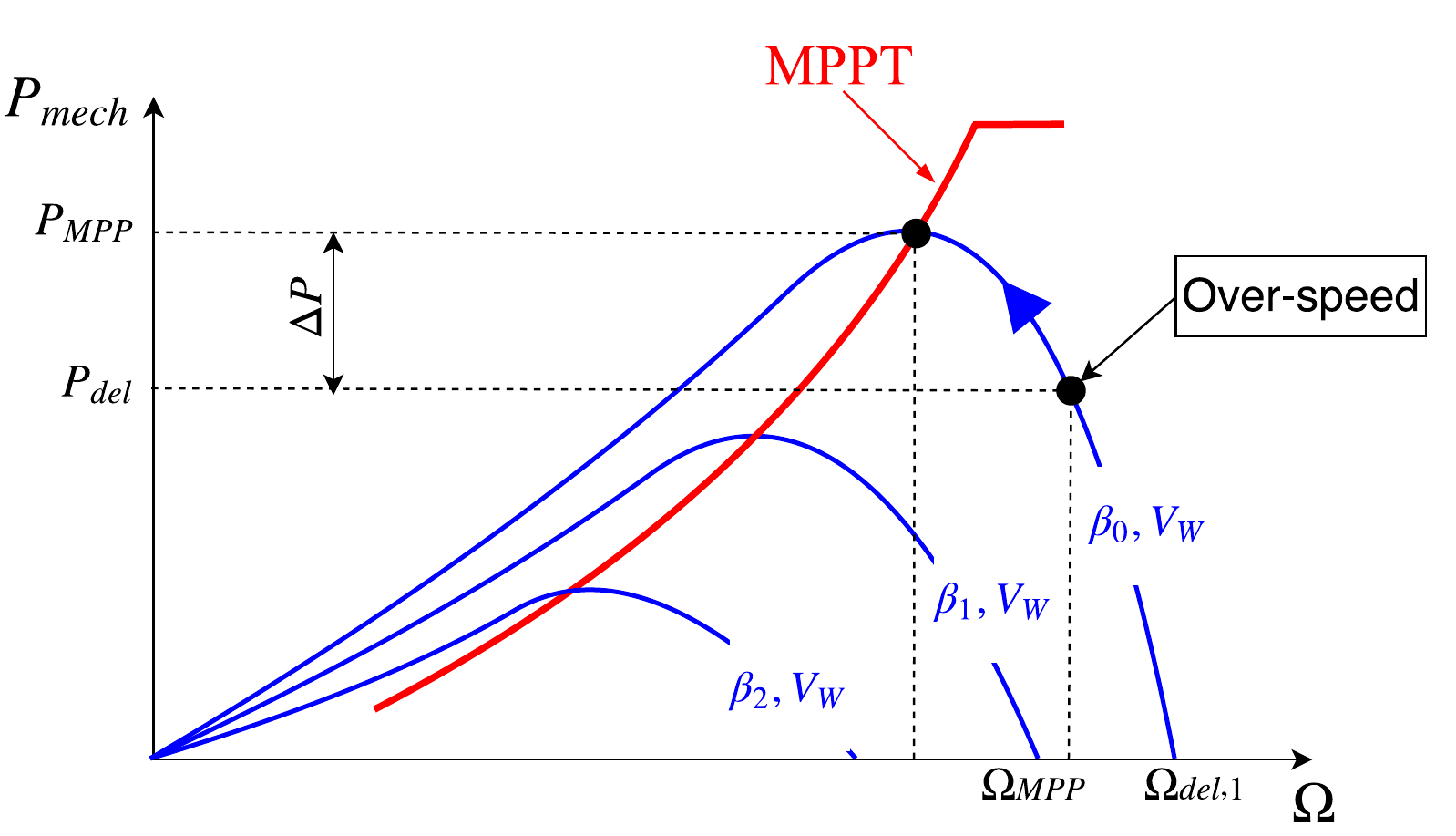}\label{fig.deload-wf1}}
	\hfill
	\subfigure[Under-speed control]{\includegraphics[width=0.79\columnwidth]{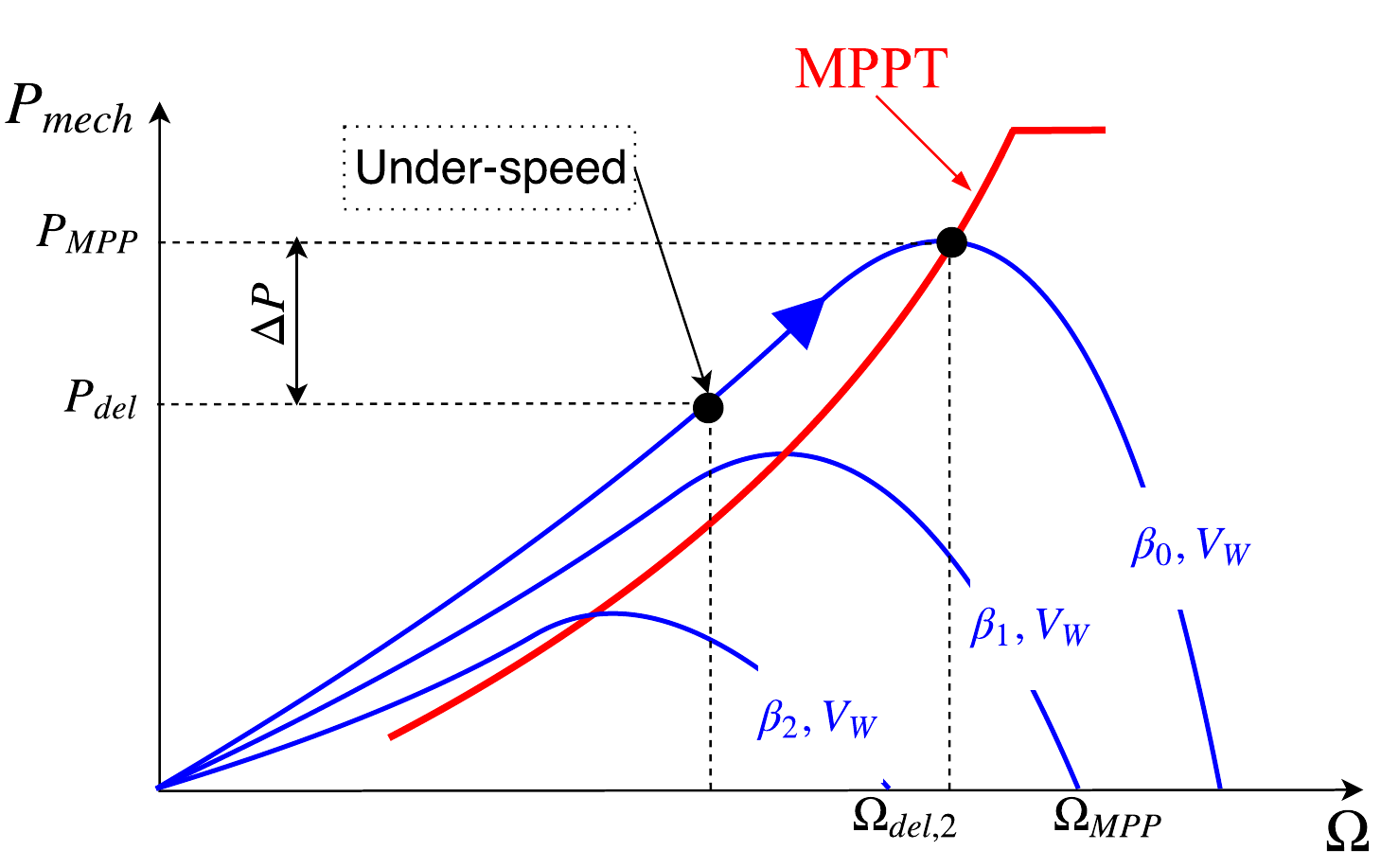}\label{fig.deload-wf21}}
	\caption{Over-speed and under-speed control}
	\label{fig.deload}
\end{figure}

Wind turbines have two possibilities to operate with the de-loading technique: $(i)$~pitch angle control and $(ii)$~over-speed control~\cite{yingcheng11}. The pitch angle control consists of increasing the pitch angle from $\beta_{0}$ to $\beta_{1}$ for a constant wind speed $V_{W}$, keeping the rotor speed at the maximum power point $\Omega_{MPP}$ (Figure~\ref{fig.deload-wf}). This way, the power supplied $P_{del}$ is below the maximum available aerodynamic power $P_{MPP}$. Therefore, a certain amount of active power reserve is available to supply additional generation in case of a frequency deviation occurs~\cite{moutis09,ma10,moutis12,vzertek12}. The over-speed control shifts the de-loaded power $P_{del}$ towards the right of the maximum power $P_{MPP}$, maintaining the pitch angle $\beta_{0}$ for a constant wind speed $V_{W}$, see Figure~\ref{fig.deload-wf1}. When frequency response is provided, rotor speed has to be reduced from $\Omega_{del,1}$ to $\Omega_{MPP}$, releasing kinetic energy to the system~\cite{castro12,vidyanandan13,wang18,zhang18}. As depicted in Figure~\ref{fig.deload-wf21}, a third possibility could be to set the turbine to operate the rotor speed below the rotor speed for MPPT operation. 
In that case, the rotor speed must increase from $\Omega_{del,2}$ to $\Omega_{MPP}$ utilizing some power extracted from the turbine. As a consequence, the frequency response is reduced, and could even be opposite to the desired behavior during the first seconds. Because of this, it is usually considered as a `detrimental strategy'~\cite{janssens07,ramtharan07}.

With regard to providing an inertial response from wind power plants, the main idea is to increase the output power of the VSWT for a few seconds. One or more supplementary loops are introduced into the active power control, which are only activated under frequency deviations. Both blades and rotor inertia are then used to provide primary frequency response under power imbalance situations. The kinetic energy stored in the rotating masses is supplied to the grid as an additional active power~\cite{alsharafi18}. 

\begin{figure}[tbp]
	\centering
	\subfigure[Droop characteristic]{\includegraphics[width=0.9\columnwidth]{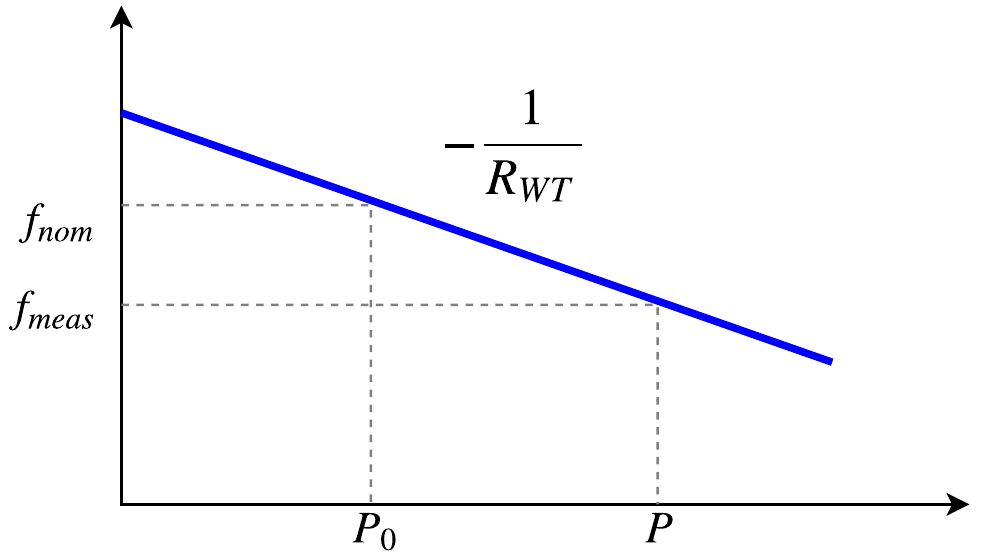}\label{fig.droop1}}
	\hfil
	\subfigure[Block diagram of droop control~\cite{ye16}]{\includegraphics[width=0.9\columnwidth]{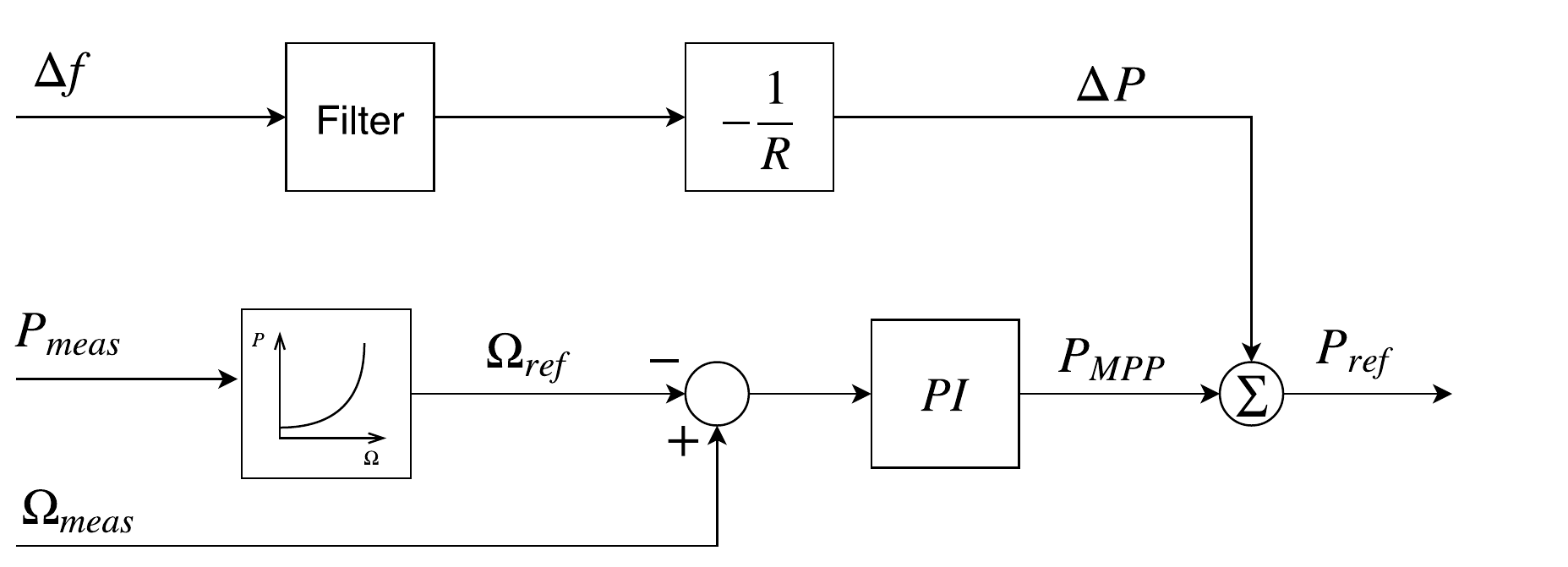}}
	\caption{Droop control for VSWTs}
	\label{fig.droop}
\end{figure}

The droop control emulates the behavior of a governor in a conventional synchronous generator, responding to the changes in the system frequency. The active power supplied by the VSWTs changes proportionally to the frequency deviation $\Delta f$ as 
illustrated in Figure~\ref{fig.droop1}, where $R_{WT}$ is the droop control setting (speed adjustment rate). Subsequently, the variation of power is defined as Eq.~\eqref{eq.pdroop}, where $\Delta P$ is the signal given to the power converter to release the stored kinetic energy. The increase of the active power output results in a decrease in the rotor speed~\cite{arani17,lertapanon17,huang17,deepak17}.
\begin{equation}
\Delta P=-\dfrac{\Delta f}{R_{WT}}
\label{eq.pdroop}
\end{equation}

\begin{figure}[tbp]
	\centering
	\subfigure[One loop]{\includegraphics[width=0.9\columnwidth]{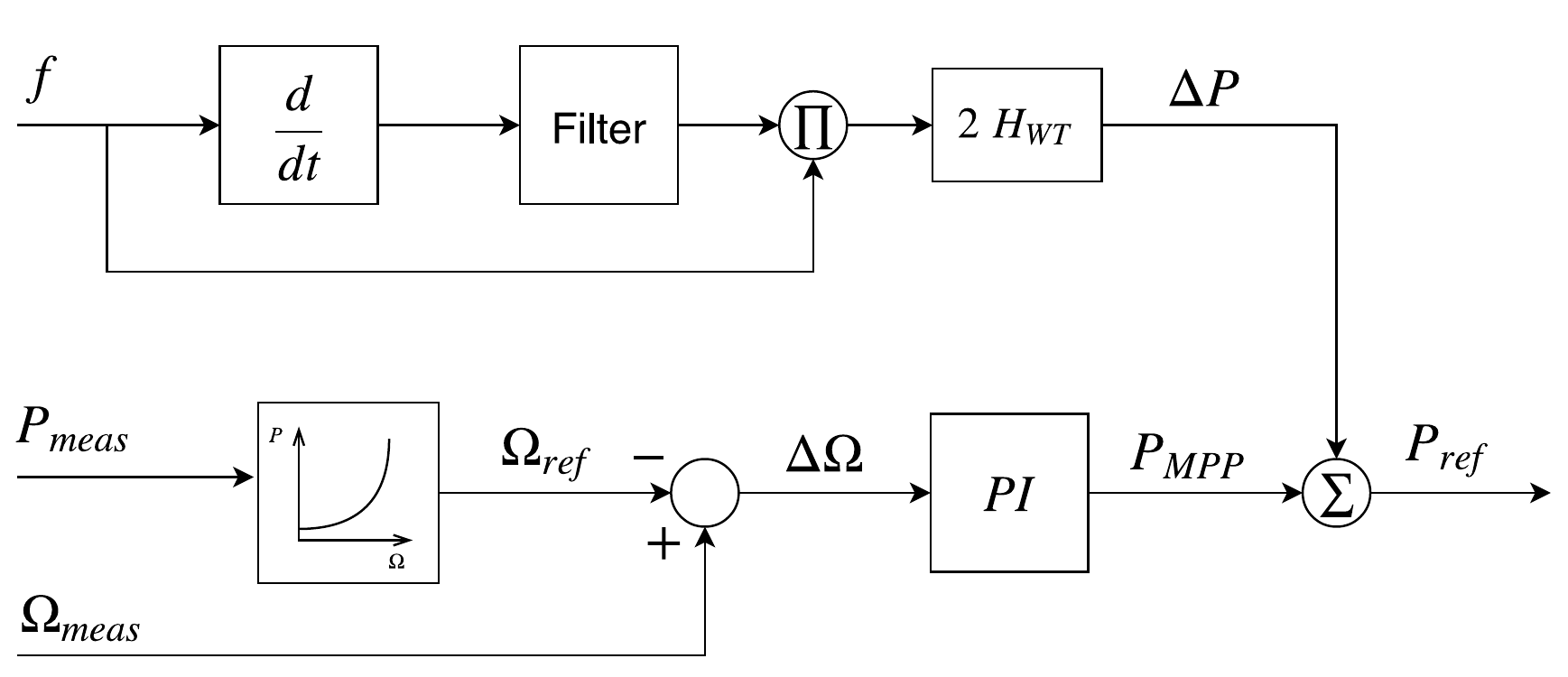}\label{fig.hidden1}}
	\hfil
	\subfigure[Two loops]{\includegraphics[width=0.9\columnwidth]{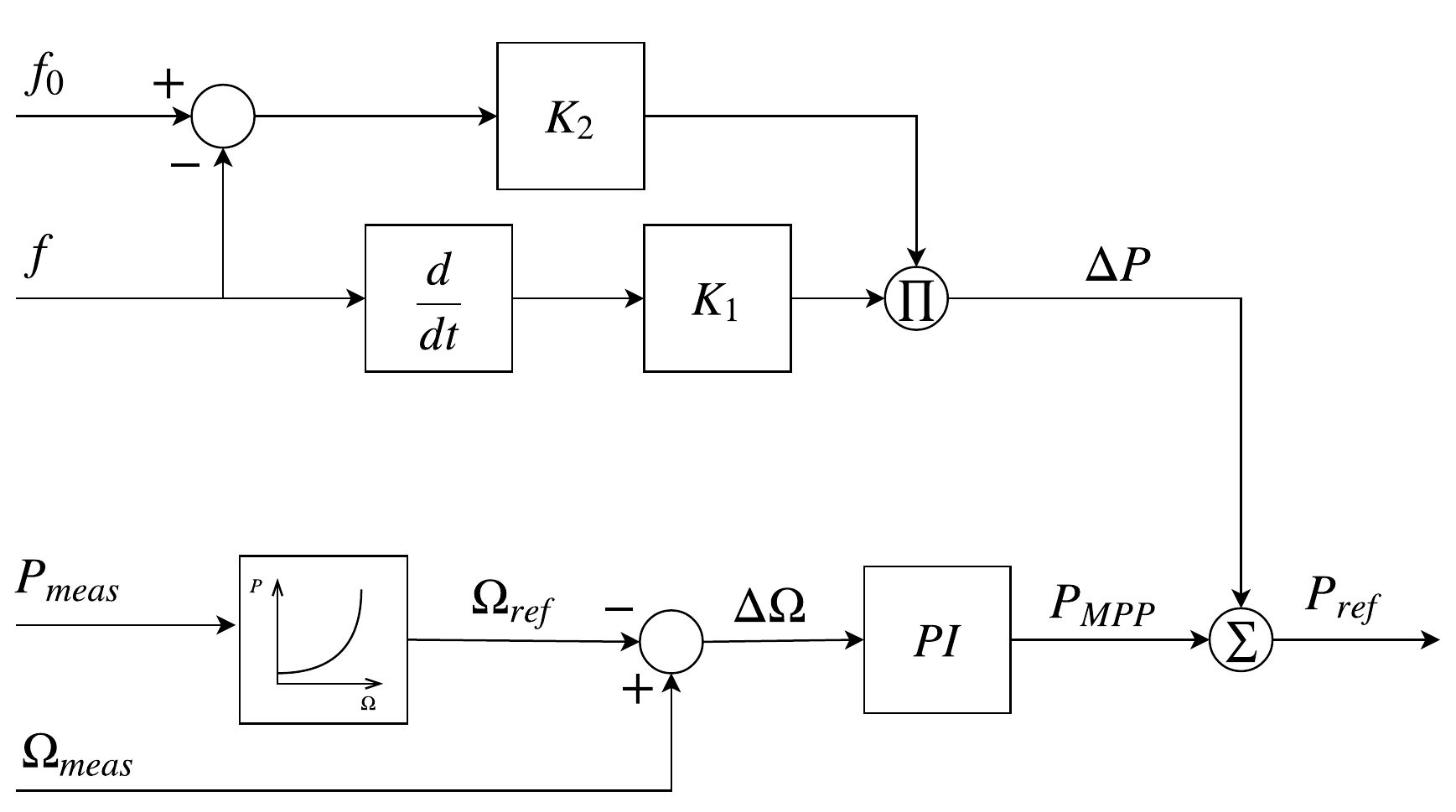}\label{fig.hidden2}}
	\caption{Hidden inertia emulation controllers}
	\label{fig.hidden}
\end{figure}

\begin{figure}[tbp]
	\centering
	\includegraphics[width=0.75\columnwidth]{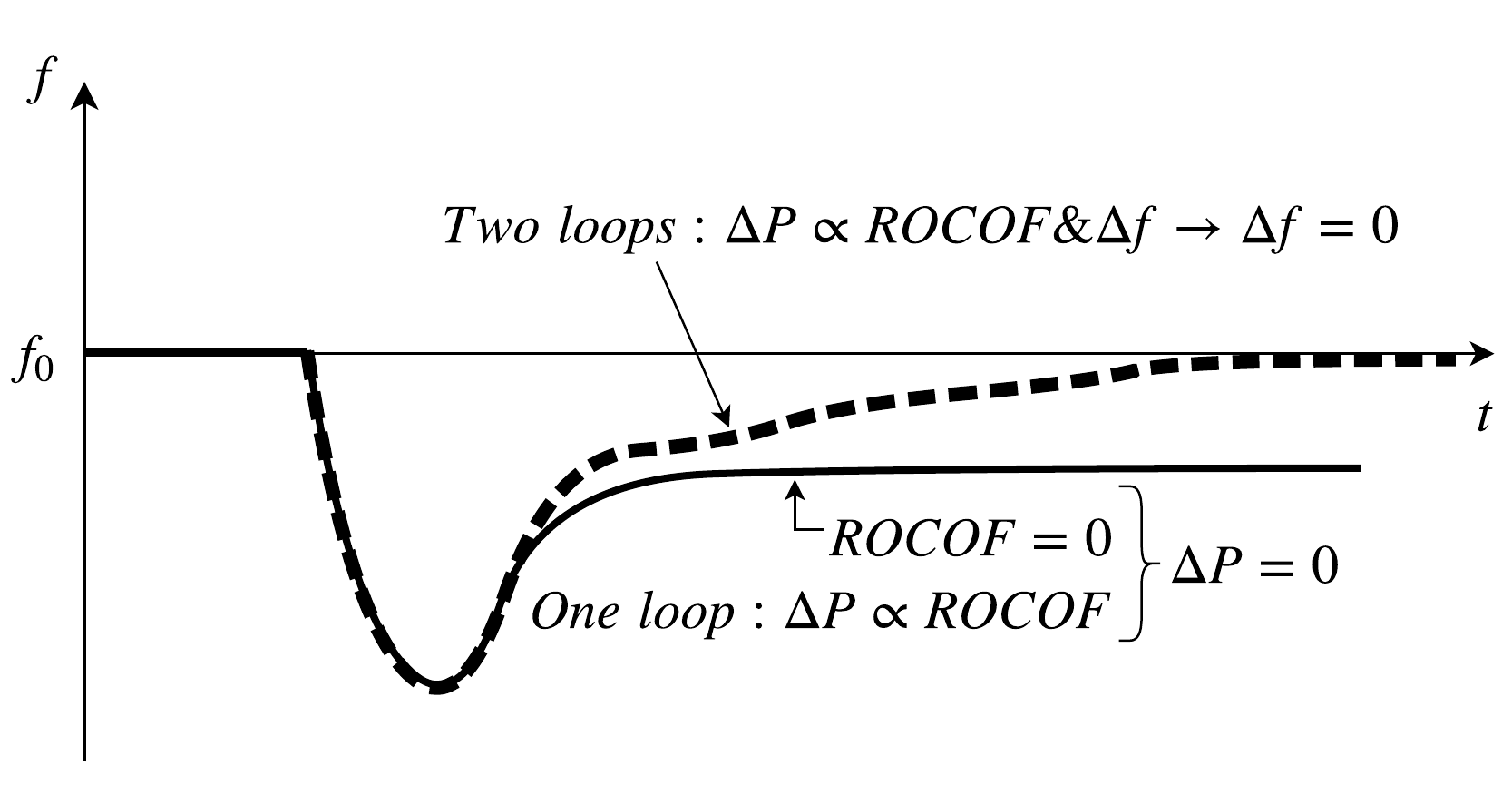}
	\caption{Frequency response of the one loop and two loops controllers}
	\label{fig.frequency_hidden}
\end{figure}

Hidden inertia emulation for wind turbines is characterized by an emulation of the inertial response of a traditional synchronous generator. There are two types of hidden inertia emulation controls: $(i)$~one loop and $(ii)$~two loops. In the first case, an additional power $\Delta P$ based on the ROCOF is added to $P_{MPP}$ after a generation deficit, thus, reducing the generator speed and releasing the stored kinetic energy of the rotating blades~\cite{gonzalez12,bonfiglio18,liu18}. The drawback of this control strategy is that frequency is not restored to its nominal value~\cite{dreidy17}. An additional loop proportional to the frequency deviation $\Delta f$ is then added, as indicated in Figure~\ref{fig.hidden2}. This second loop lasts until the frequency is restored to $f_{0}$~\cite{morren06,diaz14}. Figure~\ref{fig.frequency_hidden} compares the frequency responses by considering one or two loops controllers.

\begin{figure}[tbp]
	\centering
	\subfigure[$P-\Omega$ curve]{\includegraphics[width=0.9\columnwidth]{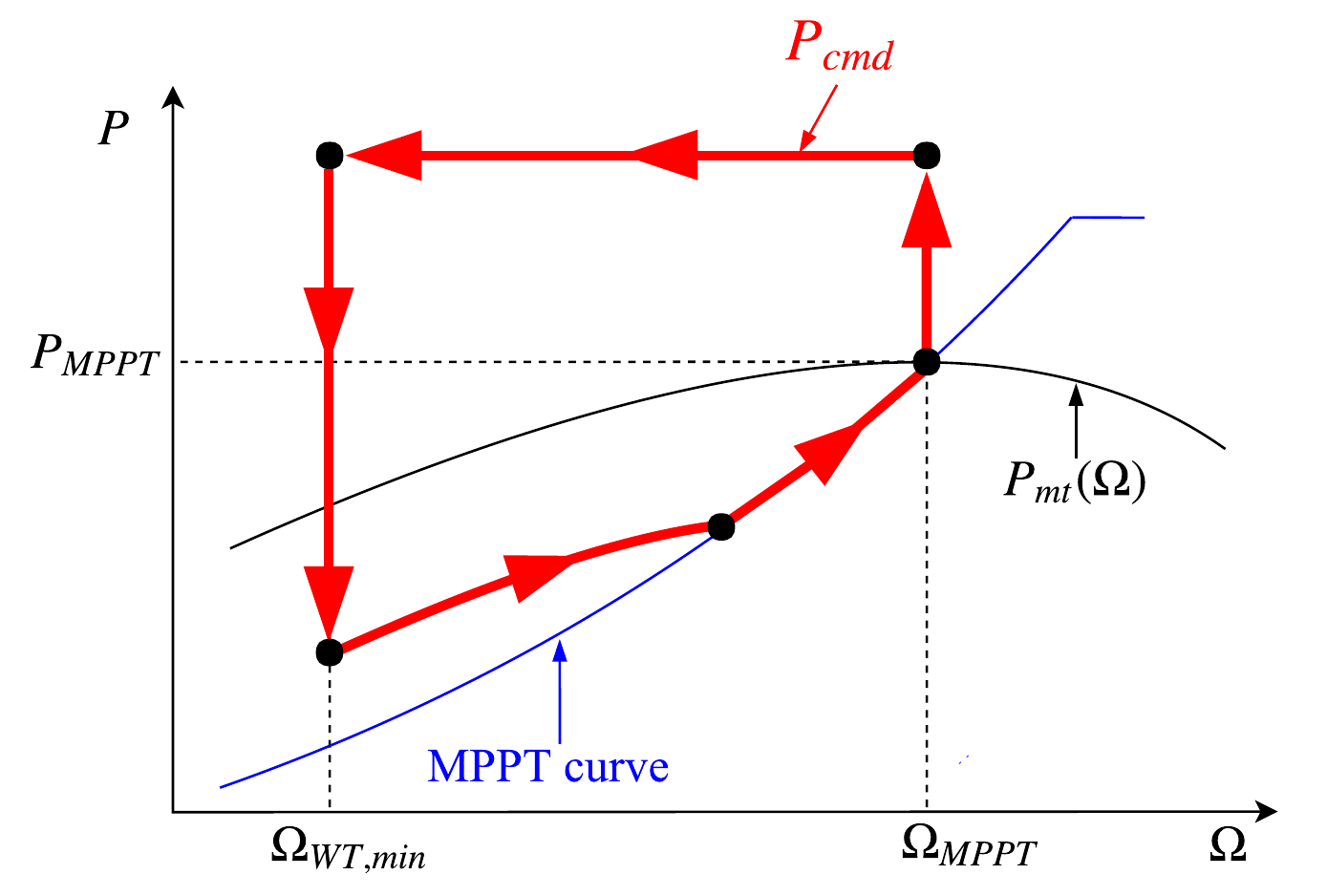}\label{fig.control_org}}
	\hfil
	\subfigure[Power variation]{\includegraphics[width=0.9\columnwidth]{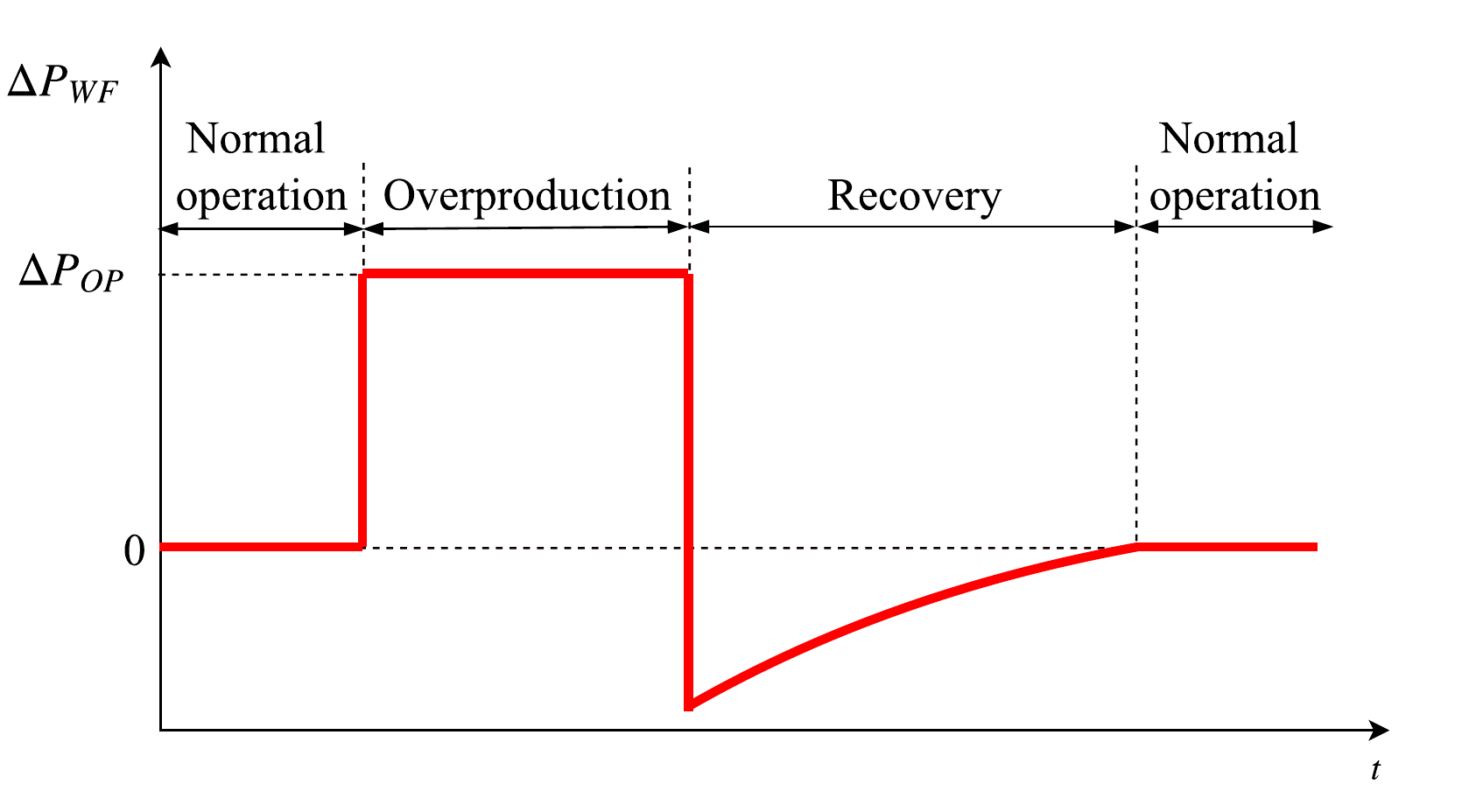}\label{fig.power_tarn}}
	\caption{Fast power reserve emulation technique \cite{tarnowski09}}
	\label{fig.fast}
\end{figure}

The fast power reserve technique is based on supplying the kinetic energy stored in the rotating masses of the wind turbine to the grid as additional active power. 
Afterward, the energy extracted is recovered through an under-production period. When the frequency deviation surpasses the predefined threshold value, the additional active power is provided, decreasing the rotational speed of the rotor. Overproduction power was initially defined as a constant value~\cite{ullah08,tarnowski09,keung09,el11,hansen14,hafiz15}. However, new approaches consider it as variable~\cite{kang16,fernandez18,fernandez18fast} by considering other limits (e.g. toque limit, the current limit of the power electronic switches, etc). The recovery period is used to restore both power and rotational speed to their pre-event values. Different techniques have also been proposed in the references listed. Figure~\ref{fig.fast} shows the fast power reserve emulation control indicated in~\cite{tarnowski09}.

Table~\ref{tab.freq_wt} presents an overview of the application of some of the techniques. 
It includes the integration of wind power plants (WPP) and the power imbalance $\Delta P$; 
both in the percentage of the total capacity of the system. As can be seen, some strategies are combined, in order to improve the frequency deviation after the generation-load mismatch.
\begin{table}[tbp]
	\begin{center}		
		\resizebox{\columnwidth}{!}{
		\begin{tabular}{ c c c c c } \hline
			\bfseries Ref. & \textbf{Type of control} &\bfseries WPP (\%) & \textbf{$\Delta P$} (\%) & \textbf{Year}\\ \hline
			\cite{margaris12} & Droop & 46 & 14 & 2012 \\ 
			\cite{margaris12}& Hidden inertia $(i)$ & 46 & 14  & 2012 \\ 
			\cite{margaris12}& Droop + Hidden inertia $(i)$ & 46 & 14  & 2012 \\ 
			\cite{chang11}& Variable droop & 30  & --  & 2011 \\ 
			\cite{wilches16} & De-loading by pitch & 24 & 3 & 2016 \\ 
			\cite{wilches16} & De-loading by pitch & 50 &  4  & 2016 \\ 
			\cite{zhang17}& Fast power reserve & 57 & 8.5  & 2017 \\ 
			\cite{su12}& Hidden inertia $(ii)$ & 25  &  1.7 & 2012 \\ 
			\cite{hwang16}& Dynamic droop + Hidden inertia $(i)$ & 10 &  8.5, 10, 11 & 2016 \\ 
			\cite{van16}& Droop + Hidden inertia $(i)$ & 15  & 2 & 2016 \\ 
			\cite{van16}& Droop + Hidden inertia $(i)$ & 50  & 2 & 2016 \\ 
			\cite{kang15}& Fast power reserve & 12.5 & 6.25 & 2015 \\ 
			\cite{you15}& Hidden inertia $(i)$ & 20 & 8.33 & 2015 \\ 
			\cite{you15}& Droop & 20 &  8.33 & 2015 \\ 
			\cite{you15}& Droop & 20 &  8.33 & 2015 \\ 
			\cite{zhang13} & Hidden inertia $(ii)$ & 30 &  2.5 & 2013\\ 
			\cite{zhang12} & Hidden inertia $(i)$ & 38 & 2.3 & 2012\\ 
			\cite{zhang12} & De-loading by pitch + Over-speed & 38 & 2.3 & 2012\\ 
			\cite{zhang12} & Hidden inertia $(i)$ + Pitch + Over-speed & 38  & 2.3 & 2012\\\hline
		\end{tabular}
	}
		\caption{Wind turbines frequency control proposals}
		\label{tab.freq_wt}
	\end{center}
\end{table}

\section{Conclusion}\label{sec:conclusion} {\color{black}An extensive
  literature review focused on inertia estimation for power systems
  and wind power plants is conducted by the authors. The contribution
  of PV power plants as a 'virtual inertia' is also discussed in the
  paper, as well as a detailed analysis of the damping factor
  evolution. Averaged inertia values are estimated for different
  regions and countries for the last two decades. Conventional
  generation units are considered accordingly, summarizing their
  inertia constant values in accordance with each type of technology and
  rated power. Our findings indicate that, nowadays, Europe presents a
  significant averaged inertia decreasing --around 20\% in the last
  two decades--, mainly due to the renewable integration decoupled
  from the grid --from 14\% in 1996 to 31\% in 2016--. With regard to
  wind turbines, they present inertia values similar to conventional
  generation units --between 2 and 6~s depending on technologies--,
  which is commonly considered as 'emulated hidden inertia'. The paper
  provides significant information for wind turbines frequency control
  strategies and studies of current power systems with high renewable
  energy source integration.}

\section*{Funding}
This work was supported by the Spanish Education, Culture and Sports Ministry [FPU16/04282].


\bibliography{biblio}

\end{document}